# Collective All-Carbon Magnetism in Triangulene Dimers

Shantanu Mishra[1]†, Doreen Beyer[2]†, Kristjan Eimre[1], Ricardo Ortiz[3,4], Joaquín Fernández-Rossier[5], Reinhard Berger[2], Oliver Gröning[1], Carlo A. Pignedoli[1], Roman Fasel[1,6], Xinliang Feng[2]*, and Pascal Ruffieux[1]*

[1]nanotech@surfaces Laboratory, Empa—Swiss Federal Laboratories for Materials Science and Technology, 8600 Dübendorf, Switzerland

[2]Center for Advancing Electronics and Department of Chemistry and Food Chemistry, Technical University of Dresden, 01062 Dresden, Germany

[3]Department of Applied Physics, University of Alicante, 03690 Sant Vicent del Raspeig, Spain

[4]Department of Chemical Physics, University of Alicante, 03690 Sant Vicent del Raspeig, Spain

[5]QuantaLab, International Iberian Nanotechnology Laboratory, 4715-330 Braga, Portugal

[6]Department of Chemistry and Biochemistry, University of Bern, 3012 Bern, Switzerland

†These authors contributed equally to this work

*E-mail: xinliang.feng@tu-dresden.de, pascal.ruffieux@empa.ch

**Abstract:** Triangular zigzag nanographenes, such as triangulene and its π-extended homologues, have received widespread attention as organic nanomagnets for molecular spintronics, and may serve as building blocks for high-spin networks with long-range magnetic order—of immense fundamental and technological relevance. As a first step toward these lines, we present the on-surface synthesis and a proof-of-principle experimental study of magnetism in covalently bonded triangulene dimers. On-surface reactions of rationally-designed precursor molecules on Au(111) lead to the selective formation of triangulene dimers in which the triangulene units are either directly connected through their minority sublattice atoms, or are separated via a 1,4-phenylene spacer. The chemical structures of the dimers have been characterized by bond-resolved scanning tunneling microscopy. Scanning tunneling spectroscopy and inelastic electron tunneling spectroscopy measurements reveal collective singlet-triplet spin excitations in the dimers, demonstrating efficient inter-triangulene magnetic coupling.

**Introduction**

The fusion of benzenoid rings in a triangular fashion leads to the generation of triangular zigzag nanographenes (TZNGs) for which no Kekulé valence structures can be drawn without leaving unpaired electrons.[1] The underlying basis for the non-Kekulé structure of TZNGs is an inherent sublattice imbalance in the bipartite honeycomb lattice such that the simultaneous pairing of all $p_z$-electrons into π-bonds is impossible (Scheme 1).[2–4] Application of Ovchinnikov's rule[5,6] predicts an increasing ground state total spin quantum number $S$ with increasing size of TZNGs. Derivatives of phenalenyl[7] (three fused rings, $S = 1/2$) and triangulene[8,9] (six fused rings, $S = 1$) have been obtained in solution, and their magnetic ground states have been confirmed by electron paramagnetic resonance spectroscopy. In the last three years, unsubstituted triangulene[10] and its larger homologues,[11,12] that is, π-extended [4]- and [5]-triangulene containing ten and fifteen fused rings, with $S = 3/2$ and 2, respectively, have been obtained on metal and insulator surfaces, and their electronic structures have been elucidated at submolecular resolution using scanning tunneling microscopy and spectroscopy (STM and STS). A range of applications have been envisaged for TZNGs in molecular electronics and spintronics such as spin filters,[13,14] qubits for quantum information processing[15] and electrically-controllable magnetic switches.[16,17] Given their high-spin ground states, interesting fundamental and technological prospects lie in the construction of one-dimensional chains and two-dimensional networks incorporating TZNGs as building blocks—such as the discovery of elusive quantum states of matter[18] and room temperature long-range magnetic ordering.[19–21] With the advent of on-surface synthesis as a chemical toolbox,[22] fabrication of extended TZNG nanostructures seems feasible on metal surfaces, given the proper chemical precursor design. Scheme 1 illustrates the versatility of TZNG nanostructures. Connecting two triangulene units directly through their minority sublattice carbon atoms does not produce a net sublattice imbalance in the structure, and is thus expected to yield an $S = 0$ ground state as per Ovchinnikov's rule, which could either correspond to an open-shell singlet or a non-magnetic, closed-shell ground state. Introduction of an organic spacer in the structure serves to not only tune the magnetic coupling between the triangulene units, but also modify the magnetic correlations, leading to high- or low-spin ground states. As shown in Scheme 1, while separation of two triangulene units by a 1,4-phenylene spacer is expected to result in an $S = 0$ ground state, separation through a 1,3-phenylene spacer generates a net sublattice imbalance in the structure, and therefore should result in an $S > 0$ ground state. Therefore, a range of nanoarchitectures based on TZNGs can be conceived with tunable coupling strengths and magnetic ground states.

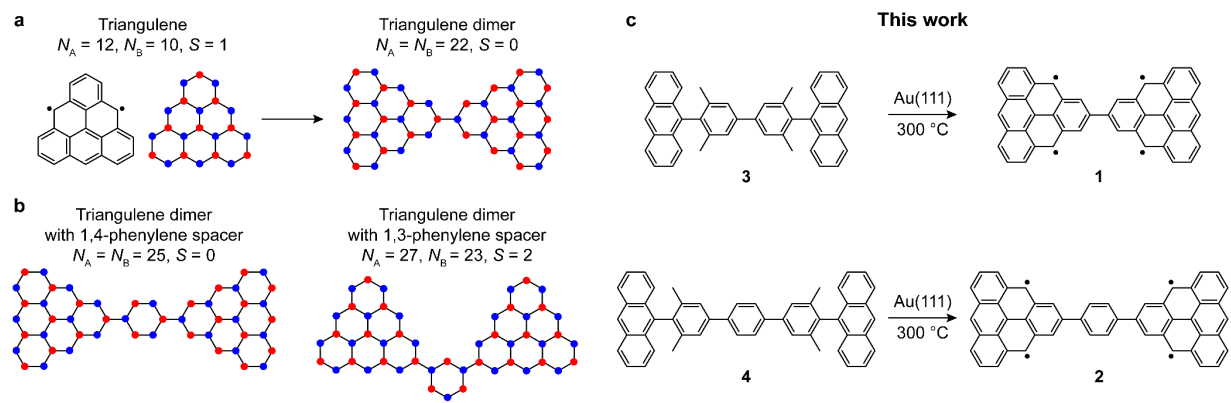

**Scheme 1.** Tunability of magnetic coupling and synthetic route toward triangulene dimers. a) Chemical structure of triangulene with the carbon atoms of the two interpenetrating triangular sublattices highlighted with blue and red filled circles (left). $N_A$ and $N_B$ denote the number of carbon atoms in the $A$ and $B$ sublattices, respectively. Triangulene exhibits a sublattice imbalance of two, with the majority sublattice atoms located at the zigzag edges. Direct coupling of two triangulene units through their minority sublattice atoms leads to no sublattice imbalance in the dimer (right). b) Schematic showing triangulene dimers with a 1,4-phenylene (left) and 1,3-phenylene (right) spacers. The dimer with 1,3-phenylene spacer contains a net sublattice imbalance of four in the structure. c) Synthetic route toward triangulene dimers reported in this work.

In this regard, two fundamental problems need to be solved. First, a direct proof of magnetism in TZNGs on metal surfaces, such as spin excitations or Kondo interactions between unpaired spins and conduction electrons of surfaces, is lacking.[23–26] Current experimental interpretation of magnetism in TZNGs is indirect, which relies on (1) spectroscopic detection of the spin-split frontier molecular orbitals, and (2) subsequent comparison of the experimental Coulomb gap with theoretical predictions to estimate the magnetic ground state. Second, it is imperative to demonstrate that spins in TZNG nanostructures can couple on a metal surface to result in a measurable collective magnetic ground state. Here, we devise a strategy to address the above problems through on-surface synthesis of triangulene dimers where the constituent triangulene units are either directly connected via a carbon-carbon bond through their minority sublattice atoms (**1**), or are separated via a 1,4-phenylene spacer (**2**). Our synthetic strategy relies on the solution synthesis of precursor molecules 9,9'-(3,3',5,5'-tetramethyl-[1,1'-biphenyl]-4,4'-diyl)dianthracene (**3**) and 9,9'-(3,3'',5,5''-tetramethyl-[1,1':4',1''-terphenyl]-4,4''-diyl)dianthracene (**4**) (Scheme 1, see Supporting Information for solution synthesis data), which, when annealed on a Au(111) surface, yield **1** and **2**, respectively. Using STS and STM-based inelastic electron tunneling spectroscopy (IETS), we unravel unambiguous spectroscopic signatures of collective magnetism in **1** and **2** in the form of singlet-triplet spin excitations.

**Results and Discussion**

Toward the synthesis of **1**, a submonolayer coverage of **3** was deposited on a Au(111) surface held at room temperature, and annealed to 300 °C to promote oxidative cyclization of the methyl groups. STM imaging of the surface after the annealing step revealed isolated dumbbell-shaped molecules and covalently bonded oligomers (Figure 1a). Figure 1b presents a high-resolution STM image of an individual molecule, which shows characteristic lobed signatures in the local density of states (LDOS). We conducted ultrahigh-resolution STM imaging with a carbon monoxide-functionalized tip[27,28] to obtain the bond-resolved structure of the molecule, which confirmed the successful formation of **1** (Figure 1c and Supporting Information, Figure S1). The synthesis of **2** was conducted in a similar manner. STM imaging after a 300 °C annealing step of a Au(111) surface with pre-deposited **4** revealed isolated molecules similar in appearance to **1** (Figures 1d,e), and ultrahigh-resolution STM imaging confirmed the successful formation of **2** (Figure 1f).

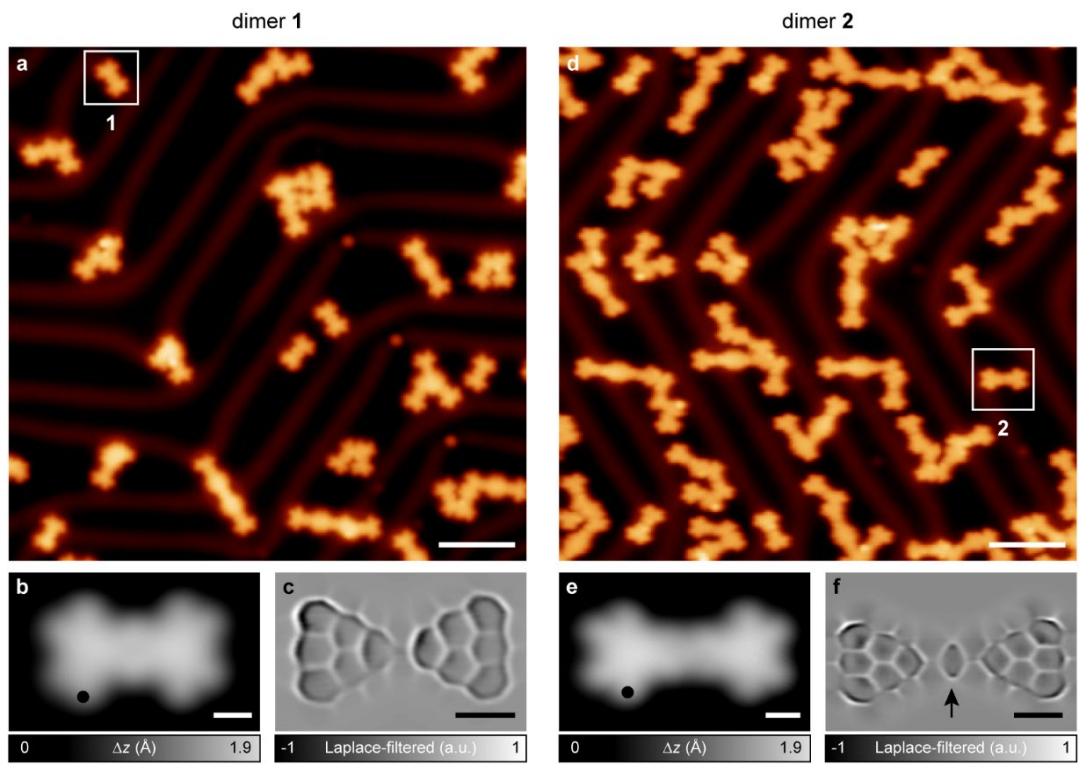

**Figure 1.** On-surface synthesis and structural characterization of **1** and **2**. a,d) Overview STM topography images after annealing precursors **3** (a) and **4** (d) on Au(111) at 300 °C. Tunneling parameters: $V = -600$ mV, $I = 100$ pA (a) and $V = -600$ mV, $I = 20$ pA (d). Isolated **1** and **2** molecules are highlighted with squares. b,e) High-resolution STM images of **1** (b) and **2** (e). Tunneling parameters: $V = -600$ mV, $I = 200$ pA (b) and $V = -600$ mV, $I = 150$ pA (e). The filled circles indicate the positions where the corresponding d$I$/d$V$ and/or IETS spectra shown in Figure 3 were acquired. c,f) Corresponding Laplace-filtered ultrahigh-resolution STM images of **1** (c) and **2** (f). The arrow in (f) highlights the 1,4-phenylene spacer. Open feedback parameters: $V = -5$ mV, $I = 50$ pA; $\Delta z = -80$ pm (c) and $-92$ pm (f). Scale bars: 5 nm (a,d) and 0.5 nm (b,c,e,f).

Figure 2 shows the electronic and magnetic structures of triangulene and the dimers **1** and **2** at successively more refined levels of theory. We start by analyzing the three systems in the nearest neighbor tight-binding (TB) model, which disregards any electron-electron interaction. The salient features in the TB energy spectra correspond to two and four non-bonding zero-energy states (ZESs) for triangulene[2,29] and **1** (and **2**, not shown), respectively (Figure 2a). The ZESs of individual triangulene units survive in the dimers **1** and **2** given that the bridging carbon-carbon bond of **1** (and the benzenoid ring of **2**) connect minority sublattice sites of the triangulene units where the ZESs have zero amplitude (Scheme 1). Inclusion of electron-electron correlations within the mean-field Hubbard (MFH) model lifts the degeneracy of the ZESs in **1** and **2**, leading to the formation of singly occupied and singly unoccupied molecular orbitals (SOMOs and SUMOs), along with the opening up of a sizeable Coulomb gap (Figure 2a). The lowest-energy MFH solution corresponds to an antiferromagnetic order between the triangulene units of **1** and **2**, leading to an $S = 0$ open-shell singlet ground state, in agreement with Ovchinnikov's rule. In the case of a single triangulene molecule, the magnetic ground state has been found to be an open-shell triplet ($S = 1$), which is approximately 500 meV lower in energy than the closed-shell first excited state.[30] Accordingly, **1** and **2** may be considered as weakly-coupled Heisenberg spin-1 dimers, since the effective exchange coupling *between* the triangulene units, $J_{eff}$, can be assumed to be much smaller than the strong ferromagnetic coupling *within* the triangulene units, $J_{FM} < 0$ (Figure 2b). Analytical solution of the Heisenberg dimer model (Supporting Information, Note S1) for an antiferromagnetic coupling $J_{eff} > 0$ predicts an open-shell singlet ground state, with the open-shell triplet state at energy $J_{eff}$ as the first and the open-shell quintet ($S = 2$) state at energy $3J_{eff}$ as the second excited state, as shown schematically in Figure 2b. To obtain quantitative values of $J_{eff}$, we solve the Hubbard model for **1** and **2** using the exact diagonalization in the complete active space (CAS) formed by six electrons in six single-particle states—that is, the four non-bonding states, along with the HOMO−1 and LUMO+1 states, where HOMO and LUMO refer to the highest occupied and the lowest unoccupied molecular orbitals, respectively (see Supporting Information for method details). The Hubbard model is known to give results in line with those of advanced quantum chemistry methods.[30] The exact diagonalization of CAS(6,6) for both **1** and **2** yields an open-shell singlet ground state, followed by the open-shell triplet and quintet states as the first and second excited states, respectively. The energies of the states are related as $E(S = 2) − E(S = 0) = 3 \times [E(S = 1) − E(S = 0)]$ for both **1** and **2**, thus conforming to the effective Heisenberg dimer model of two antiferromagnetically coupled spin-1 systems. The magnetic excitation spectra of **1** and **2** calculated in the CAS(6,6) model, which approximate the singlet-triplet and singlet-quintet gaps, are shown in Figure 2c as a function of the on-site Coulomb repulsion $U$. Our calculations show that at a given $U$, the excitation energies are much larger for **1** than for **2**. Furthermore, $J_{eff}$ values for both **1** and **2** are at least thirty times smaller than the intra-triangulene exchange coupling $J_{FM}$, confirming the basic assumption $|J_{eff}| \ll |J_{FM}|$ underlying the Heisenberg dimer model.

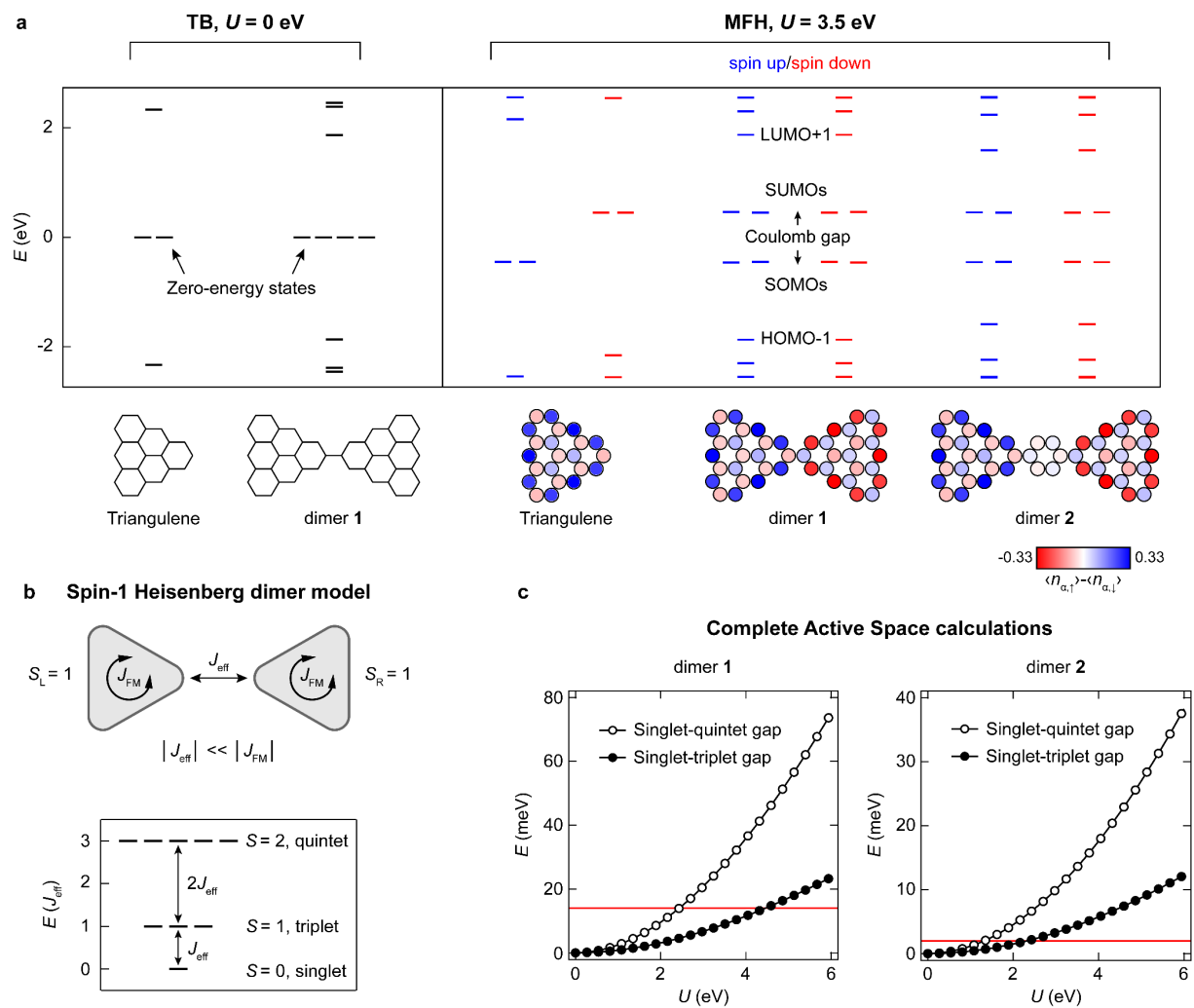

**Figure 2.** Theoretical electronic and magnetic characterization of **1** and **2**. a) Nearest neighbor TB energy spectra of triangulene and **1** (left panel) and MFH energy spectra of triangulene, **1** and **2** along with the corresponding spin polarization plots (right panel). $U$ denotes the on-site Coulomb repulsion. b) Schematic illustration of the spin-1 Heisenberg dimer model for **1** and **2** (upper panel), with the corresponding energy level scheme from an analytical solution of the Heisenberg dimer model for an antiferromagnetic coupling between the triangulene units (lower panel). $S_{L/R}$ denotes the total spin quantum numbers of the left/right triangulene units. c) Energies of the open-shell triplet and quintet states of **1** and **2** with respect to their open-shell singlet ground states calculated in the CAS(6,6) approximation, and plotted as a function of $U$. The red solid lines indicate the experimental singlet-triplet gaps of 14 meV and 2 meV for **1** and **2**, respectively.

The predicted outcomes of the theoretical analyses are convincingly demonstrated in our experiments. d$I$/d$V$ spectroscopy (where $I$ and $V$ are current and voltage, respectively) on **1** and **2** reveal broad peaks centered at ~ −400 mV and +1.25 V (Figures 3a,d). d$I$/d$V$ maps acquired at these biases exhibit close correspondence with the meanfield Hubbard local density of states (MFH-LDOS) maps of the SOMOs and SUMOs of **1** and **2** (Figures 3b,e). This confirms the detection of the spin-split frontier molecular orbitals of both species, and their Coulomb gaps approximately equal 1.65 eV. d$I$/d$V$ spectroscopy on **1** in the vicinity of the Fermi energy reveals conductance steps symmetric around zero bias (Figure 3c, blue curve), which is indicative of an inelastic excitation.[31] Given the open-shell singlet ground state and the open-shell triplet first excited state of **1**, we ascribe the inelastic excitations to singlet-triplet ($S = 0$ to $S = 1$) spin excitation, which obeys the IETS spin selection rule that dictates $\Delta S = 0, \pm 1$ for magnetic excitations (Supporting Information, Note S2). The excitation threshold, extracted from a fit to the experimental IETS spectrum with an antiferromagnetic spin-1 Heisenberg dimer model,[32] is ±14 mV, and provides a direct experimental measure of the $J_{\text{eff}}$ (or, the singlet-triplet gap) of **1** (Figure 3c, red curve and Supporting Information, Figure S2). Similarly, d$I$/d$V$ spectroscopy on **2** also presents singlet-triplet spin excitations (Figure 3f), albeit with a substantially reduced excitation threshold of ±2 mV—demonstrating the tunability of inter-triangulene magnetic coupling. The experimentally observed singlet-triplet gaps of **1** and **2** are in good agreement with CAS(6,6) calculations at reasonable values of $U$ (Figure 2c). The absence of inelastic excitations in linearly fused oligomers of **1**, where the triangulene units are separated over a large distance and therefore exhibit negligible overlap of their wave functions, further strengthens the case for the observed inelastic excitations as being spin excitations (Supporting Information, Figure S3).

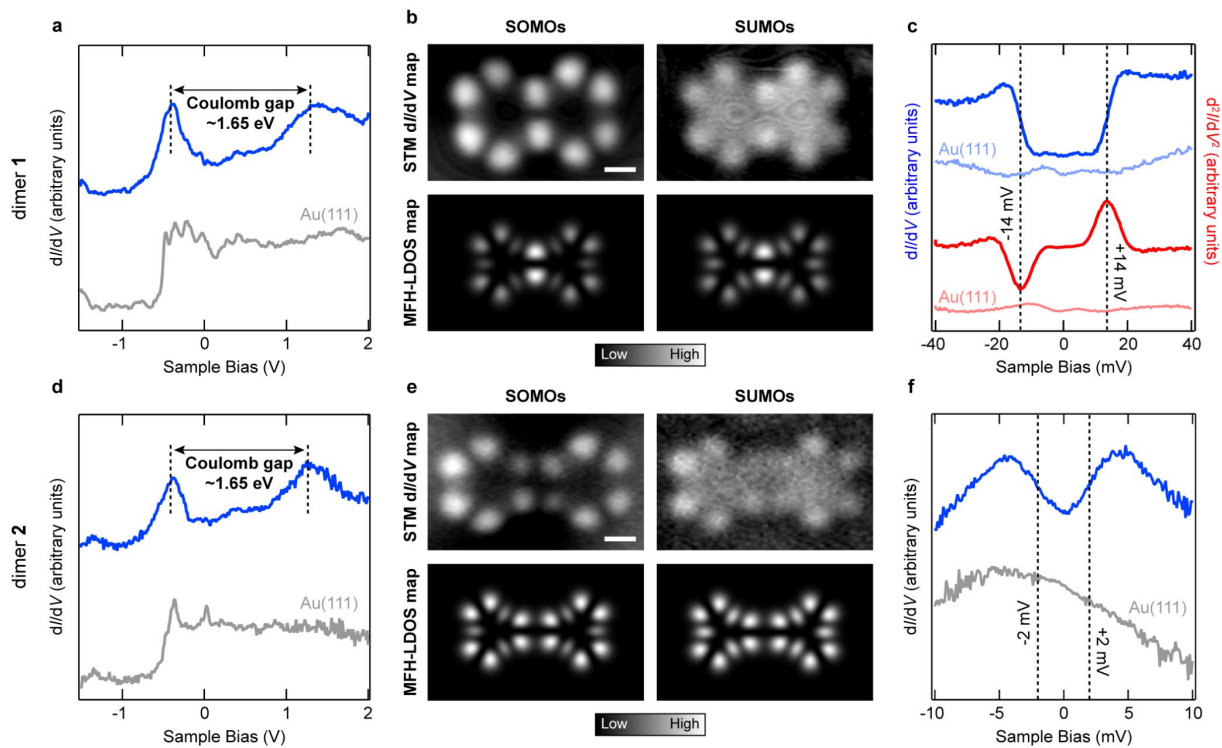

**Figure 3.** Experimental electronic and magnetic characterization of **1** and **2**. a,d) Long-range d$I$/d$V$ spectrum (blue curves) acquired on **1** (a) and **2** (d). Open feedback parameters: $V = -1.50$ V, $I = 350$ pA (a) and $V = -1.50$ V, $I = 400$ pA (d); $V_{\text{rms}} = 16$ mV (a) and 10 mV (d). b,e) Experimental d$I$/d$V$ maps (upper panels) and MFH-LDOS maps (lower panels) at the SOMOs and SUMOs resonances of **1** (b) and **2** (e). Tunneling parameters: $V = -450$ mV, $I = 350$ pA (SOMOs, b), $V = +1.00$ V, $I = 350$ pA (SUMOs, b), $V = -400$ mV, $I = 350$ pA (SOMOs, e) and $V = +1.10$ V, $I = 450$ pA (SUMOs, e); $V_{\text{rms}} = 22$ mV. c,f) d$I$/d$V$ (blue curve) and IETS (red curve) spectra acquired on **1** (c), and d$I$/d$V$ spectrum (blue curve) acquired on **2** (f) in the vicinity of the Fermi energy. Acquisition positions for the spectra are indicated by filled circles in Figures 1b,e. Open feedback parameters: $V = -40$ mV, $I = 500$ pA (d$I$/d$V$ spectra, c), $V = -40$ mV, $I = 1.2$ nA (IETS spectra, c) and $V = -10$ mV, $I = 750$ pA (d$I$/d$V$ spectra, f); $V_{\text{rms}} = 400$ μV (d$I$/d$V$ spectra) and 4 mV (IETS spectra). Scale bars: 0.5 nm.

**Conclusion**

In summary, we have demonstrated the on-surface synthesis of triangulene dimers with and without a 1,4-phenylene spacer. The magnetic ground states of both dimers are predicted to be the open-shell singlet, with the first and second excited states being the open-shell triplet and quintet, respectively. In accordance with theoretical predictions, we experimentally detect singlet-triplet spin excitations, whose strength can be tuned with the spatial separation between the triangulene units. Our results prove that TZNGs on metal surfaces retain their high-spin magnetic ground states, and can efficiently couple to give rise to collective magnetism. Given the large exchange interaction of 14 meV and the presumably small magnetic anisotropy in triangulene dimers due to the weak spin-orbit coupling in carbon, our findings should pave the way for fabrication of magnetic TZNG networks, providing a platform to explore emergent quantum phases and realize technologically relevant magnetic materials.

**Acknowledgements**

This work was supported by the Swiss National Science Foundation (grant numbers 200020-182015 and IZLCZ2-170184), the NCCR MARVEL funded by the Swiss National Science Foundation (grant number 51NF40-182892), the European Union's Horizon 2020 research and innovation program (grant number 785219, Graphene Flagship Core 2), the Office of Naval Research (grant number N00014-18-1-2708), ERC Consolidator grant (T2DCP, grant number 819698), the German Research Foundation (DFG) within the Cluster of Excellence Center for Advancing Electronics Dresden (cfaed) and EnhanceNano (grant number 391979941), the European Social Fund and the Federal State of Saxony (ESF-Project GRAPHD, TU Dresden), Generalitat Valenciana and Fondo Social Europeo (grant number ACIF/2018/175), MINECO-Spain (grant number MAT2016-78625) and the Portuguese FCT (grant number UTAPEXPL/ NTec/0046/2017).

**Conflict of Interest**

The authors declare no conflict of interest.

**Keywords:** magnetism, nanographenes, on-surface synthesis, scanning probe microscopy, surface chemistry

**References**

[1] Y. Morita, S. Suzuki, K. Sato, T. Takui, *Nat. Chem.* **2011**, *3*, 197–204.

[2] J. Fernández-Rossier, J. J. Palacios, *Phys. Rev. Lett.* **2007**, *99*, 177204.

[3] W. L. Wang, O. V. Yazyev, S. Meng, E. Kaxiras, *Phys. Rev. Lett.* **2009**, *102*, 157201.

[4] Z. Bullard, E. Costa Girão, C. Daniels, B. G. Sumpter, V. Meunier, *Phys. Rev. B* **2014**, *89*, 245425.

[5] A. A. Ovchinnikov, *Theoret. Chim. Acta* **1978**, *47*, 297–304.

[6] E. H. Lieb, *Phys. Rev. Lett.* **1989**, *62*, 1201–1204.

[7] K. Goto, T. Kubo, K. Yamamoto, K. Nakasuji, K. Sato, D. Shiomi, T. Takui, M. Kubota, T. Kobayashi, K. Yakusi, et al., *J. Am. Chem. Soc.* **1999**, *121*, 1619–1620.

[8] G. Allinson, R. J. Bushby, J. L. Paillaud, D. Oduwole, K. Sales, *J. Am. Chem. Soc.* **1993**, *115*, 2062–2064.

[9] J. Inoue, K. Fukui, T. Kubo, S. Nakazawa, K. Sato, D. Shiomi, Y. Morita, K. Yamamoto, T. Takui, K. Nakasuji, *J. Am. Chem. Soc.* **2001**, *123*, 12702–12703.

[10] N. Pavliček, A. Mistry, Z. Majzik, N. Moll, G. Meyer, D. J. Fox, L. Gross, *Nat. Nanotechnol.* **2017**, *12*, 308–311.

[11] S. Mishra, D. Beyer, K. Eimre, J. Liu, R. Berger, O. Gröning, C. A. Pignedoli, K. Müllen, R. Fasel, X. Feng, et al., *J. Am. Chem. Soc.* **2019**, *141*, 10621–10625.

[12] J. Su, M. Telychko, P. Hu, G. Macam, P. Mutombo, H. Zhang, Y. Bao, F. Cheng, Z.-Q. Huang, Z. Qiu, et al., *Sci. Adv.* **2019**, *5*, eaav7717.

[13] W. L. Wang, S. Meng, E. Kaxiras, *Nano Lett.* **2008**, *8*, 241–245.

[14] M. Ezawa, *Phys. Rev. B* **2008**, *77*, 155411.

[15] A. D. Güçlü, P. Potasz, P. Hawrylak, *Phys. Rev. B* **2011**, *84*, 035425.

[16] A. D. Güçlü, P. Potasz, O. Voznyy, M. Korkusinski, P. Hawrylak, *Phys. Rev. Lett.* **2009**, *103*, 246805.

[17] W.-L. Ma, S.-S. Li, *Phys. Rev. B* **2012**, *86*, 045449.

[18] D.-J. Choi, N. Lorente, J. Wiebe, K. von Bergmann, A. F. Otte, A. J. Heinrich, *Rev. Mod. Phys.* **2019**, *91*, 041001.

[19] G. Trinquier, N. Suaud, N. Guihéry, J.-P. Malrieu, *ChemPhysChem* **2011**, *12*, 3020–3036.

[20] X. Li, J. Zhou, Q. Wang, X. Chen, Y. Kawazoe, P. Jena, *New J. Phys.* **2012**, *14*, 033043.

[21] X. Li, Q. Wang, *Phys. Chem. Chem. Phys.* **2012**, *14*, 2065–2069.

[22] S. Clair, D. G. de Oteyza, *Chem. Rev.* **2019**, *119*, 4717–4776.

[23] C. F. Hirjibehedin, C. P. Lutz, A. J. Heinrich, *Science* **2006**, *312*, 1021–1024.

[24] J. Li, S. Sanz, M. Corso, D. J. Choi, D. Peña, T. Frederiksen, J. I. Pascual, *Nat. Commun.* **2019**, *10*, 200.

[25] S. Mishra, D. Beyer, K. Eimre, S. Kezilebieke, R. Berger, O. Gröning, C. A. Pignedoli, K. Müllen, P. Liljeroth, P. Ruffieux, et al., *Nat. Nanotechnol.* **2020**, *15*, 22–28.

[26] S. Mishra, D. Beyer, R. Berger, J. Liu, O. Gröning, J. I. Urgel, K. Müllen, P. Ruffieux, X. Feng, R. Fasel, *J. Am. Chem. Soc.* **2020**, *142*, 1147–1152.

[27] L. Gross, F. Mohn, N. Moll, P. Liljeroth, G. Meyer, *Science* **2009**, *325*, 1110–1114.

[28] G. Kichin, C. Weiss, C. Wagner, F. S. Tautz, R. Temirov, *J. Am. Chem. Soc.* **2011**, *133*, 16847–16851.

[29] M. Ezawa, *Phys. Rev. B* **2007**, *76*, 245415.

[30] R. Ortiz, R. A. Boto, N. García-Martínez, J. C. Sancho-García, M. Melle-Franco, J. Fernández-Rossier, *Nano Lett.* **2019**, *19*, 5991–5997.

[31] S. K. Khanna, J. Lambe, *Science* **1983**, *220*, 1345–1351.

[32] M. Ternes, *New J. Phys.* **2015**, *17*, 063016.

**Table of Contents graphical abstract:**

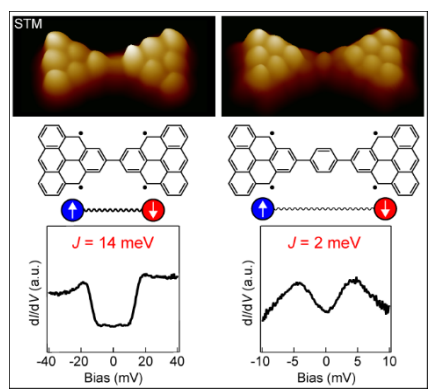

On-surface synthesis of covalently bonded triangulene dimers with or without a 1,4-phenylene spacer is achieved on Au(111). Scanning tunneling spectroscopy measurements reveal collective magnetism in the dimers in the form of singlet-triplet spin excitations, demonstrating efficient and tunable inter-triangulene magnetic coupling.

# Supporting Information

## Collective All-Carbon Magnetism in Triangulene Dimers

Shantanu Mishra, Doreen Beyer, Kristjan Eimre, Ricardo Ortiz, Joaquín Fernández-Rossier, Reinhard Berger, Oliver Gröning, Carlo A. Pignedoli, Roman Fasel, Xinliang Feng*, and Pascal Ruffieux*

# 1. Materials and methods

**1.1 Sample preparation and STM/STS measurements.** STM measurements were performed with a Scienta Omicron low-temperature LT-STM operating at 4.5 K and base pressure below $5\times10^{-11}$ mbar. Au(111) single crystal surfaces were prepared by $Ar^+$ sputtering and annealing cycles. Precursor molecules **3** and **4** were contained in quartz crucibles and deposited at 483 and 530 K, respectively, from a home-built evaporator on Au(111) held at room temperature. STM images and d$I$/d$V$ maps were acquired in constant-current mode. Unless noted otherwise, gold-coated tungsten tips were used for imaging and spectroscopy. Indicated tunneling biases are provided with respect to the sample. d$I$/d$V$ and IETS spectra, and d$I$/d$V$ maps were acquired with a lock-in amplifier operating at a frequency of 860 Hz. Lock-in modulation voltages (root mean square amplitude, $V_{rms}$) for each measurement is provided in the respective figure captions. The fitting of d$I$/d$V$ and IETS spectra to extract the spin excitation thresholds were performed using a code developed by Markus Ternes.[1] Ultra-high resolution STM images were acquired with carbon monoxide-functionalized tips, where the molecules are scanned in a constant-height mode, and the current channel is displayed. Open feedback parameters, and subsequent tip approach distances ($\Delta z$) for each measurement is provided in the respective figure captions. Carbon monoxide molecules were deposited on Au(111) at a maximum sample temperature of 13 K. The data shown in this study were processed and analyzed with WaveMetrics Igor Pro or WSxM software.[2]

**1.2. Tight binding calculations of the electronic structure.** The tight-binding calculations of **1** and **2** have been performed by numerically solving the mean-field Hubbard Hamiltonian with nearest neighbor hopping:

$$\widehat{H}_{MFH} = -t \sum_{\langle \alpha,\beta \rangle,\sigma} c^\dagger_{\alpha,\sigma} c_{\beta,\sigma} + U \sum_{\alpha,\sigma} \langle n_{\alpha,\sigma} \rangle n_{\alpha,\bar{\sigma}} - U \sum_\alpha \langle n_{\alpha,\uparrow} \rangle \langle n_{\alpha,\downarrow} \rangle, \tag{S1}$$

Here, $c^\dagger_{\alpha,\sigma}$ and $c_{\beta,\sigma}$ denote the spin selective ($\sigma \in \{\uparrow,\downarrow\}$ with $\bar{\sigma} \in \{\downarrow,\uparrow\}$) creation and annihilation operator at neighboring sites $\alpha$ and $\beta$, $t$ is the nearest neighbor hopping parameter (with $t$ = 2.7 eV used), $U$ is the on-site Coulomb repulsion, $n_{\alpha,\sigma}$ is the number operator and $\langle n_{\alpha,\sigma} \rangle$ is the mean occupation number at site α. Orbital electron densities, $\rho$, of the $n^{th}$-eigenstate with energy $E_n$ have been simulated from the corresponding state vector $a_{n,i,\sigma}$ by:

$$\rho_{n,\sigma}(\vec{r}) = \left| \sum_i a_{n,i,\sigma} \phi_{2p_z}(\vec{r} - \vec{r}_i) \right|^2, \tag{S2}$$

where $i$ denotes the atomic site index, and $\phi_{2p_z}$ denotes the Slater $2p_z$ orbital for carbon.

**1.3. Complete Active Space (CAS) calculations.** The CAS method, described by Ortiz et al.,[3] can be broken down in the following steps:

1. Solution of the one-orbital tight-binding model for a given structure and choice of hopping parameters. This yields a single particle spectrum and a set of molecular orbitals.
2. Representation of the Hubbard model in the basis of molecular orbitals.
3. Choice of active space orbitals. In our calculations we include the four non-bonding zero-energy states and the lowest energy pair of finite-energy states above and below the non-bonding states (that is, HOMO−1 and LUMO+1).
4. Construction of the many-body configurations for six electrons in six orbitals.

    The number of configurations is $C_6(12) = \binom{12}{6} = 924$.

5. Construction of the many-body matrix Hamiltonian, obtained by the representation of the Hubbard model in this basis.

6. Diagonalization of the many-body matrix and analysis of energy spectrum degeneracies, that permit to identify the multiplets.

**1.4. Solution synthesis.** Unless otherwise noted, all starting chemical materials were purchased from Sigma Aldrich, TCI, ABCR, and other chemical providers. All starting materials were used as received without further purification. The solution chemical reactions, unless otherwise mentioned, were conducted under air- and moisture-free conditions using a sealed Schlenk system under argon atmosphere, because of handling air- and moisture-sensitive chemical substances. The reaction progress was monitored by thin layer chromatography (TLC), containing silica-coated aluminum plates and fluorescence marker $F_{254}$ (silica 60, $F_{254}$, Merck). If necessary, crude reaction products were purified by preparative silica gel chromatography (particle size: 40–63 μm, VWR Chemicals) and recycling gel permeation chromatography (*r*GPC). *r*GPC was carried out on JAI HPLC LC 9110 II NEXT instrument with fraction collector FC-3310, and in series connected GPC columns 2H and 1H with chloroform (HPLC grade) as eluent. For structural characterization, proton and carbon nuclear magnetic resonance spectra ($^1$H and $^{13}$C-NMR, respectively) were recorded at room temperature (296 K) on a BRUKER AC 300 P NMR instrument, operating at 300 MHz for $^1$H-NMR and 75 MHz for $^{13}$C-NMR. The NMR measurements were carried out in the liquid-state using deuterated dichloromethane as solvent ($CD_2Cl_2$, 99.8 atom% D, $\delta_{\text{H-NMR}}$ = 5.32 ppm / $\delta_{\text{C-NMR}}$ = 54.2 ppm), purchased from Euroisotop. The peak pattern in $^1$H-NMR spectra is described by the commonly used abbreviations: s = singlet, d = doublet, t = triplet and m = multiplet. High-resolution matrix-assisted laser desorption/ionization time-of-flight (HR-MALDI-TOF) mass spectra (MS) were obtained in the liquid-state on Autoflex Speed MALDI-TOF instrument from BRUKER, using 1,8-dihydroxyanthrone (dithranol) and *trans*-2-[3-(4-*tert*-Butylphenyl)-2-methyl-2-propenylidene]malononitrile (DCTB) as matrix. High-resolution atmospheric pressure chemical ionization MS (HR-APCI-MS) and electrospray ionization MS (HR-ESI-MS) were recorded with the Agilent 6538 Ultra High Definition Accurate-Mass Q-TOF LC/MC system. Elemental analysis from recrystallized solid compounds was carried out on a varia MICRO cube from Elementar. The solid was burned at 1150 °C for 70 s under oxygen supply. The melting points from solid compounds were determined with the melting point M-560 instrument from BÜCHI. The temperature range was set to 90–340 °C, with a temperature interval of 10 °C/min. The measurements were performed in melting point tubes from Marienfeld (80 x 1.5 mm, one-side open), and the melting point temperature was recorded once the sample was completely melted.

## 2. Supporting STM, STS and theoretical data

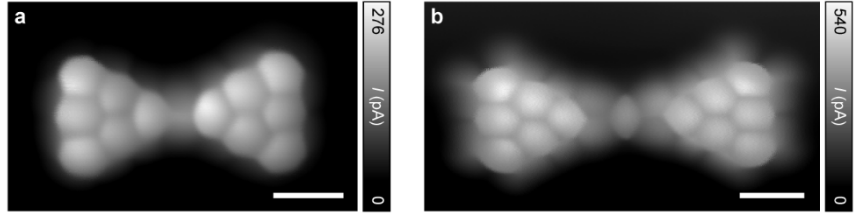

**Figure S1.** Raw data of ultrahigh-resolution STM images of **1** and **2**. a,b) Ultrahigh-resolution STM images of **1** (a) and **2** (b) (raw data). The corresponding Laplace-filtered images are shown in Figures 1c,f. Open feedback parameters: $V = -5$ mV, $I = 50$ pA; $\Delta z = -80$ pm (a) and -92 pm (b). Scale bars: 0.5 nm.

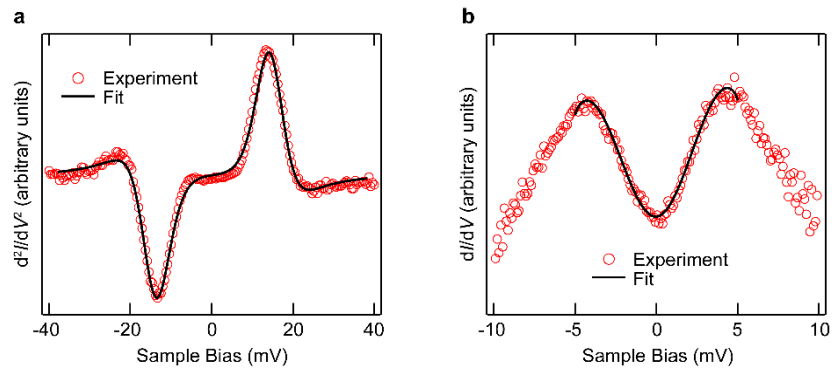

**Figure S2.** Fits to experimental spin excitation spectra. a,b) Experimental IETS spectrum acquired on **1** (a) and d$I$/d$V$ spectrum acquired on **2** (b) in the vicinity of the Fermi energy, revealing singlet-triplet spin excitations (open circles). The data in (a) and (b) are shown in Figures 3c,f. The solid curves are fit to the experimental data, from which spin excitation thresholds of ±14 mV and ±2 mV are extracted for **1** and **2**, respectively. Open feedback parameters: $V = -40$ mV, $I = 1.2$ nA (a) and $V = -10$ mV, $I = 750$ pA (b); $V_{rms} = 4$ mV (a) and $V_{rms} = 400$ µV (b).

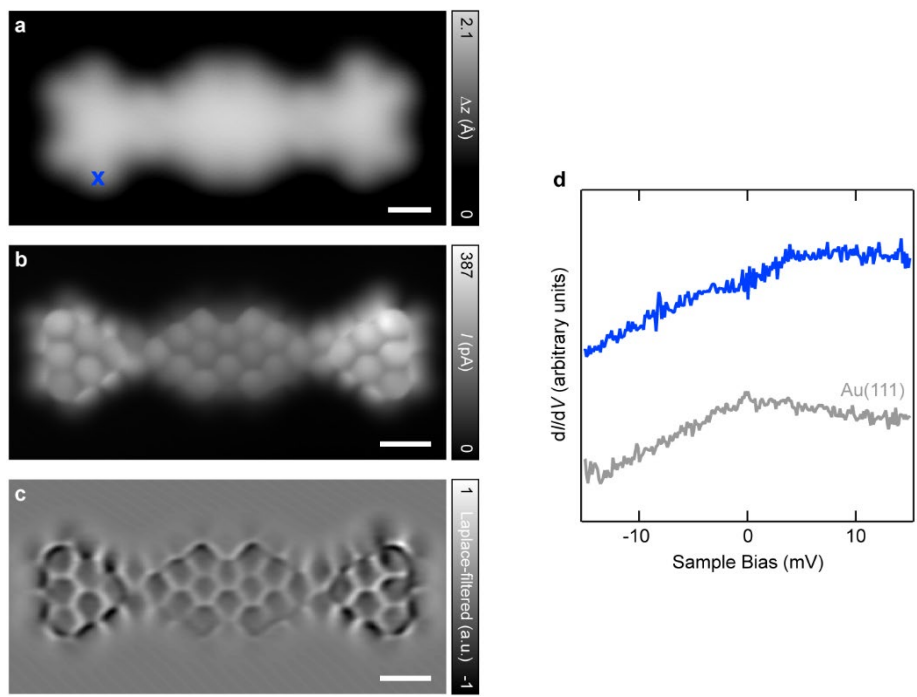

**Figure S3.** Absence of spin excitations in fused dimers of **1**. a–c) High-resolution (a), ultrahigh-resolution (b) and corresponding Laplace-filtered ultrahigh-resolution (c) STM images of a fused dimer of **1**, where the triangulenes are separated over a large distance. d) d$I$/d$V$ spectrum acquired on the fused dimer in the vicinity of the Fermi energy (blue curve) revealing absence of inelastic excitations, therefore proving the absence of magnetic coupling between the triangulenes. Acquisition position of the spectrum in (d) is highlighted by a cross in (a). Open feedback parameters: $V = -20$ mV, $I = 500$ pA; $V_{rms} = 400$ μV. Scale bars: 0.5 nm.

**Note S1: Solution of the Heisenberg dimer model.** The energy spectrum for a Heisenberg dimer Hamiltonian

$$\hat{H} = J\mathbf{S}_1 \cdot \mathbf{S}_2, \tag{S3}$$

where $J$ is the exchange coupling between the spins and $\mathbf{S}_1$ and $\mathbf{S}_2$ correspond to the individual spin operators, can be obtained using the following trick. We define the total spin operator

$$\mathbf{S} = \mathbf{S}_1 + \mathbf{S}_2. \tag{S4}$$

We use the fact that the spectrum of $S^2 = \mathbf{S} \cdot \mathbf{S}$ is $S(S+1)$, where $S$ are the integer/half integer numbers that cover the range $|S_1 - S_2|, \ldots, S_1 + S_2$. We now write

$$S^2 = (\mathbf{S}_1 + \mathbf{S}_2)^2 = S_1^2 + S_2^2 + 2\mathbf{S}_1 \cdot \mathbf{S}_2. \tag{S5}$$

The spectrum of the first two operators on the right hand side of equation (S5) is $S_{1,2}(S_{1,2}+1)$. Therefore, we can write

$$\hat{H} = J\mathbf{S}_1 \cdot \mathbf{S}_2 = \frac{J}{2}[S(S+1) - S_1(S_1+1) - S_2(S_2+1)]. \tag{S6}$$

For triangulene, $S_1 = S_2 = 1$, and $S$ can thus take three values, that is, $S=0$, $S=1$ and $S=2$. The energies of the three spin states are given by:

$$E(S) = \frac{J}{2}[S(S+1) - 4]. \tag{S7}$$

For $J > 0$ (that is, antiferromagnetic coupling between the triangulenes), the ground state has $S = 0$, and we have $E(S) - E(0) = (J/2)S(S+1)$, which yields the excitation energies

$$E(1) - E(0) = J \text{ and } E(2) - E(0) = 3J, \tag{S8}$$

as also obtained through the CAS(6,6) method.

**Note S2: IETS spin selection rule.** Here we elaborate on the origin of the $\Delta S = 0, \pm 1$ spin selection rule for IETS. Our starting point is the assumption that the inelastic co-tunneling event is a spin conserving process when both the molecule and the tunneling electron are considered. Therefore, the initial and final total spin must be conserved, that is

$$S_i(total) = S_f(total). \tag{S9}$$

Now, the initial spin state of the tunneling electron and the molecule is the one obtained from combining the initial spin of the molecule $S_i$ and the $S = 1/2$ for the electron

$$S_i(total) = S_i \pm \frac{1}{2}. \tag{S10}$$

Similarly, the final spin state is also expressed in terms of the final spin of the molecule $S_f$ and the $S = 1/2$ for the electron

$$S_f(total) = S_f \pm \frac{1}{2}. \tag{S11}$$

Now, combining equations (S9)–(S11), we arrive at the condition

$$\Delta S = S_f - S_i = 0, \pm 1. \tag{S12}$$

which provides the selection rule for observing spin excitations in IETS.

## 3. Synthetic procedures

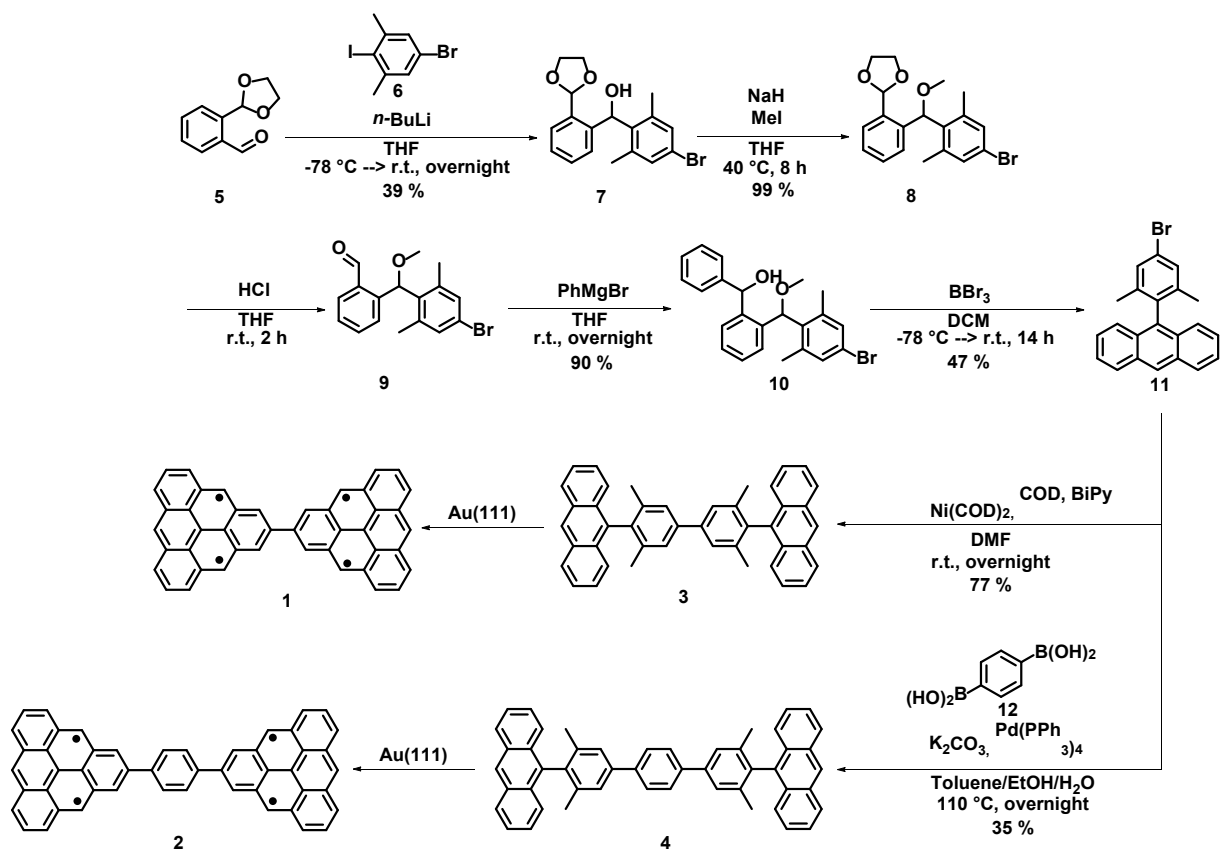

**Scheme S1.** Summary of synthetic procedures toward formation of **1** and **2**.

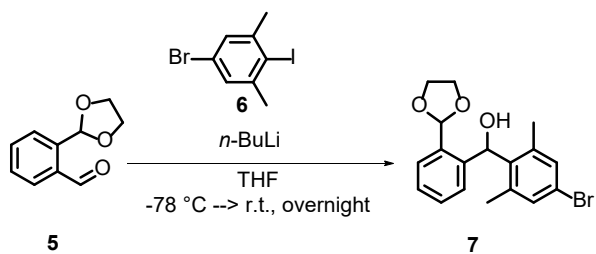

**Scheme S2.** Synthesis of compound **7**.

**(2-(1,3-Dioxolan-2-yl)phenyl(4-bromo-2,6-dimethylphenyl)methanol (7):** Commercially available 5-bromo-2-iodo-*m*-xylene (**6**) (3.5 g, 11.2 mmol, 1.0 eq.) was dissolved in 45 ml dry tetrahydrofuran (THF) and cooled to -78 °C. A solution of n-butyl lithium (*n*-BuLi) in hexane (1.6 M, 7.7 ml, 12.3 mmol, 2.2 eq.) was added dropwise under argon atmosphere and the reaction was maintained at -78 °C for 1 hour. Separately, 2-(1,3-dioxolan-2-yl)benzaldehyde (**5**) (1.0 g, 5.6 mmol, 1.0 eq.), synthesized according to literature procedure,[4] was dissolved in 5 ml dry THF and added under argon atmosphere to the reaction mixture via syringe. The resulting mixture was allowed to warm up gradually to room temperature and the reaction mixture stirred until completion. The reaction mixture was quenched with an aqueous solution of ammonium chloride (NH4Cl), extracted three times with ethyl acetate (EA), and the combined organic layer was washed with brine and dried over magnesium sulfate (MgSO4). The solvent excess was removed by evaporation and the crude compound was purified by silica gel chromatography using EA/*iso*-hexane 1:3 as eluent to afford **7** as white solid (800 mg, 39 %).

**¹H-NMR** (CD$_2$Cl$_2$, 300 MHz): δ = 7.60 (dd, *J* = 7.4, 1.6 Hz, 1H), 7.35 – 7.17 (m, 4H), 6.97 (dd, *J* = 7.6, 1.4 Hz, 1H), 6.51 (d, *J* = 3.2 Hz, 1H), 6.03 (s, 1H), 4.28 – 4.03 (m, 4H), 3.72 (d, *J* = 3.2 Hz, 1H), 2.27 (s, 6H) ppm.

**¹³C-NMR** (CD$_2$Cl$_2$, 75 MHz): δ = 141.4, 140.4, 137.6, 135.8, 132.4, 130.2, 128.6, 128.5, 128.5, 121.3, 104.0, 70.3, 66.0, 65.9, 22.0 ppm.

**HR-ESI-MS** (positive mode): calc. for [M+Na]$^+$: 385.0415, found for [M+Na]$^+$: 385.0401 (deviation: 3.63 ppm).

**Melting point**: 111.8 °C.

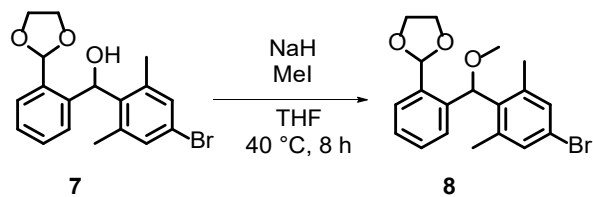

**Scheme S3.** Synthesis of compound **8**.

**2-(2-((4-Bromo-2,6-dimethylphenyl)(methoxy)methyl)phenyl)-1,3-dioxolane (8):** A suspension of sodium hydride (NaH) (171.7 mg, 7.1 mmol, 4.0 eq.) in 20 ml dry THF was heated to 40 °C. Compound **7** (650.0 mg, 1.8 mmol, 1.0 eq.), which has been dissolved in 5 ml dry THF, was slowly added under argon atmosphere. The mixture was stirred at 40 °C for 30 minutes. Afterwards, methyl iodide (MeI) (762.0 mg, 5.3 mmol, 3.0 eq.) was added in one portion. After stirring for 8 hours at 40 °C the reaction mixture was cooled to room temperature and the solvent residue was removed under reduced pressure. The crude compound was purified by silica gel chromatography using EA/*iso*-hexane 1:3 as eluent to obtain **8** as sticky light orange oil (670 mg, 99 %).

**¹H-NMR** (CD$_2$Cl$_2$, 300 MHz): δ = 7.68 (dd, *J* = 7.7, 1.4 Hz, 1H), 7.36 – 7.11 (m, 4H), 6.85 (d, *J* = 7.7 Hz, 1H), 6.15 (s, 1H), 6.12 (s, 1H), 4.22 – 3.96 (m, 4H), 3.31 (d, *J* = 4.6 Hz, 3H), 2.23 (s, 6H) ppm.

**¹³C-NMR** (CD$_2$Cl$_2$, 75 MHz): δ = 141.3, 138.5, 138.2, 135.1, 132.3, 129.2, 128.6, 128.0, 127.0, 121.7, 101.5, 79.8, 65.6, 57.2, 21.1 ppm.

**HR-APCI-MS** (positive mode): calc. for [M-OMe]$^+$: 345.0485, found for [M-OMe]$^+$: 345.0485.

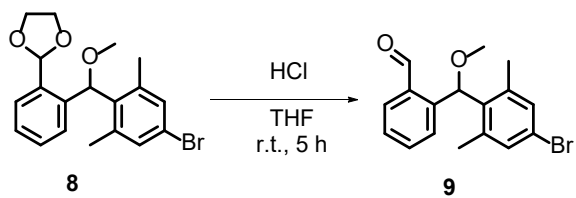

**Scheme S4.** Synthesis of compound **9**.

**2-((4-Bromo-2,6-dimethylphenyl)(methoxy)methyl)benzaldehyde (9):** Compound **8** (200.0 mg, 0.5 mmol, 1.0 eq.) was dissolved in a mixture of 6 ml THF and 6 ml 10 % aqueous hydrochloric acid (HCl). The mixture was stirred at room temperature for 5 hours and then neutralized by adding a diluted, aqueous solution of sodium bicarbonate (NaHCO$_3$). After extraction with EA three times, the combined organic phase was washed with brine and dried over MgSO$_4$. The excess of organic solvent was evaporated under reduced pressure and compound **9** was obtained as yellow sticky oil, which has been directly used for the next reaction step without further purification.

**¹H-NMR** (CD$_2$Cl$_2$, 300 MHz): δ = 10.36 (s, 1H), 7.95 – 7.81 (m, 1H), 7.42 (pd, *J* = 7.4, 4.2 Hz, 2H), 7.27 (s, 2H), 6.92 – 6.77 (m, 1H), 6.28 (s, 1H), 3.33 (s, 3H), 2.22 (s, 6H) ppm.

**¹³C-NMR** (CD$_2$Cl$_2$, 75 MHz): δ = 192.6, 141.6, 141.3, 136.3, 134.2, 133.7, 132.6, 130.3, 129.1, 128.1, 122.2, 79.0, 57.0, 21.2 ppm.

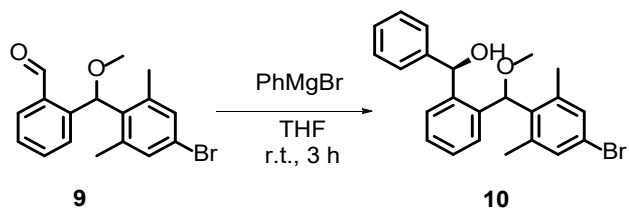

**Scheme S5.** Synthesis of compound **10**.

**(1*R*))-2-((4-Bromo-2,6-dimethylphenyl)(methoxy)methyl)phenyl)(phenyl)metha-nol (10):** Compound **9** (500.0 mg, 1.5 mmol, 1.0 eq.) has been dissolved in 5 ml dry THF. A solution of phenylmagnesium bromide (PhMgBr) in hexane (1.0 M, 3.0 ml, 0.54 mmol, 2.0 eq.) was added dropwise and the mixture was stirred at room temperature for 3 hours. After quenching with an aqueous solution of NH$_4$Cl the mixture was extracted three times with EA, washed with brine and dried over MgSO$_4$. Target compound **10** was isolated as light colorless sticky oil after flash silica gel chromatography using EA/*iso*-hexane 1:3 as eluent (555 mg, 90 %).

**$^1$H-NMR** (CD$_2$Cl$_2$, 300 MHz): $\delta$ = 7.41 – 7.24 (m, 11H), 6.71 (d, *J* = 7.8 Hz, 1H), 6.28 (d, *J* = 2.8 Hz, 1H), 5.95 (s, 1H), 3.34 (s, 3H), 2.14 (s, 6H) ppm.

**$^{13}$C-NMR** (CD$_2$Cl$_2$, 75 MHz): $\delta$ = 145.5, 143.3, 141.1, 136.7, 134.3, 132.6, 129.3, 129.0, 128.3, 128.1, 128.0, 128.0, 81.0, 72.4, 56.9, 21.1 ppm.

**HR-ESI-MS** (positive mode): calc. for [M+Na]$^+$: 433.0779, found for [M+Na]$^+$: 433.0769 (deviation: 2.31 ppm).

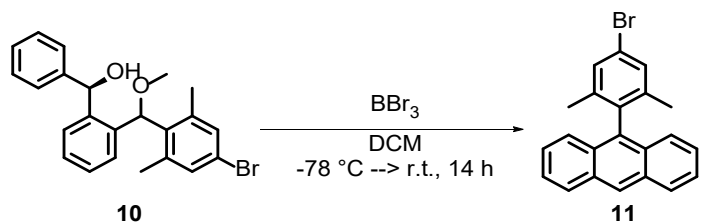

**Scheme S6.** Synthesis of compound **11**.

**9-(4-Bromo-2,6-dimethylphenyl)anthracene (11):** Intermediate compound **10** (600.0 mg, 1.5 mmol, 1.0 eq.) was dissolved in 10 ml dry dichloromethane (DCM) and cooled to -78 °C. A solution of boron tribromide (BBr$_3$) in hexane (1.0 M, 1.75 ml, 1.75 mmol, 1.2 eq.) was added dropwise under argon atmosphere and the solution was allowed to warm up gradually to room temperature. After stirring for 14 hours at room temperature the reaction was quenched with an aqueous solution of NH$_4$Cl. Afterwards, the mixture was extracted three times with DCM, washed with brine and dried over MgSO$_4$. The crude target was purified by silica gel chromatography using DCM/*iso*-hexane 1:9 as eluent to afford **11** as light yellow solid (250 mg, 47 %).

**$^1$H-NMR** (CD$_2$Cl$_2$, 300 MHz): $\delta$ = 8.54 (s, 1H), 8.09 (d, *J* = 8.5 Hz, 2H), 7.61 – 7.29 (m, 8H), 1.71 (s, 6H) ppm.

**$^{13}$C-NMR** (CD$_2$Cl$_2$, 75 MHz): $\delta$ = 140.7, 137.4, 134.6, 132.2, 130.8, 123.0, 129.3, 127.19, 126.6, 125.9, 125.9, 121.9, 20.1 ppm.

**HR-MALDI-TOF** (matrix: dithranol): calc. for [M]$^+$: 360.0513, found for [M]$^+$: 360.0565 (deviation: 14.44 ppm).

**HR-ESI-MS** (positive mode): calc. for [M]$^+$: 360.0513, found for [M]$^+$: 360.0502 (deviation: 3.05 ppm).

**Elemental analysis C$_{22}$H$_{17}$Br**: calc. for C: 73.14, H: 4.74; found for C: 73.361, H: 4.576 %.

**Melting point**: 163.9 °C

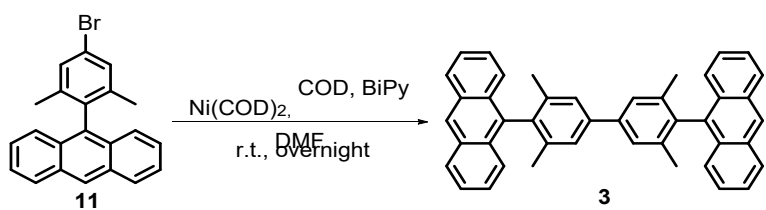

**Scheme S7.** Synthesis of compound **3**.

**9,9'-(3,3',5,5'-tetramethyl-[1,1'-biphenyl]-4,4'-diyl)dianthracene (3):** A mixture of 2,2'-bipyridine (BiPy)) (2.6 mg, 17.0 μmol, 1.2 eq.), bis(1,5-cyclooctadiene)nickel(0) (Ni(COD)$_2$) (4.5 mg, 17 μmol, 1.2 eq.) and 1,5-cyclooctadiene (COD) (1.8 mg, 17 μmol, 11.2 eq.) dissolved in 1 ml *N,N*-dimethylformamide (DMF) has been prepared and stirred at room temperature for 30 minutes under glove box conditions. Afterwards, 9-(4-bromo-2,6-dimethylphenyl)anthracene (**11**) (10.0 mg, 28.0 μmol, 2.0 eq.) was added in one portion and the mixture was stirred at room temperature overnight. Afterwards, the reaction mixture was quenched with water and was extracted three times with DCM. The combined organic layer was washed with brine and dried over MgSO$_4$. After removing the solvent excess under reduced pressure, the crude compound was purified by silica gel chromatography using DCM/*iso*-hexane 1:9 as eluent to afford title compound **3** as light yellow solid (6 mg, 77 %).

**$^1$H-NMR** (CD$_2$Cl$_2$, 300 MHz): $\delta$ = 8.56 (s, 2H), 8.12 (d, *J* = 8.5 Hz, 4H), 7.70 (s, 4H), 7.61 – 7.48 (m, 8H), 7.45 – 7.36 (m, 4H), 1.85 (s, 12H) ppm.

**$^{13}$C-NMR** (CD$_2$Cl$_2$, 75 MHz): $\delta$ = 140.9, 138.8, 137.2, 136.0, 132.3, 130.2, 129.2, 126.8, 126.7, 126.4, 125.9, 20.7.

**HR-MALDI-TOF** (matrix: DCTB): calc. for [M]$^+$: 562.2660, found for [M]$^+$: 562.2674 (deviation: 2.49 ppm).

**HR-APCI-MS** (positive mode): calc. for [M+H]$^+$: 563.2739, found for [M+H]$^+$: 563.2728 (deviation: 1.95 ppm).

**Elemental analysis C$_{44}$H$_{34}$:** calc. for C: 93.91, H: 6.09; found for C: 93.81, H: 6.13 %.

**Melting point**: > 340 °C.

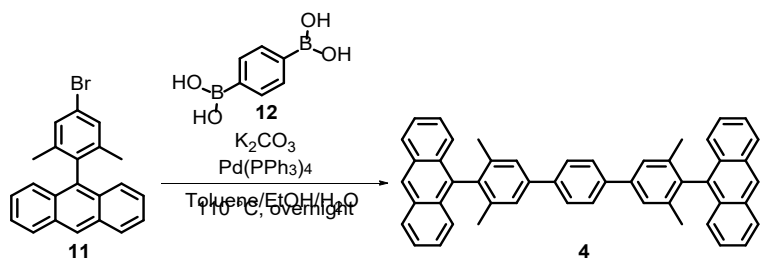

**Scheme S8.** Synthesis of compound **4**.

**9,9'-(3,3'',5,5''-tetramethyl-[1,1':4',1''-terphenyl]-4,4''-diyl)dianthracene (4):** A mixture of 3 ml ethanol (EtOH), 1 ml water (H$_2$O) and 1 ml toluene has been prepared and intensively purged with argon for at least 15 minutes. Meanwhile, compound **11** (37.0 mg, 0.2 mmol, 1.0 eq.), commercially available 1,4-phenylenediboronic acid (**12**) (201.0 mg, 0.5 mmol, 2.5 eq.) and potassium carbonate (K$_2$CO$_3$) (277.0 mg, 2.0 mmol, 9.0 eq.) have been added. At last, tetrakis(triphenylphosphine)palladium(0) (Pd(PPh$_3$)$_4$) (15.0 mg, 0.01 mmol, 0.06 eq.) was quickly added in one portion and reaction mixture was heated to 110 °C overnight. After cooling down to room temperature the reaction mixture was extracted three times with DCM, washed with brine and dried over MgSO$_4$. The crude material was purified by silica column chromatography using DCM/*iso*-hexane 1:9 as eluent to afford **4** as dark yellow solid (50 mg, 35 %). Further purification by *r*GPC afford **4** as yellow solid.

**$^1$H-NMR** (CD$_2$Cl$_2$, 300 MHz): $\delta$ = 8.56 (s, 2H), 8.12 (d, *J* = 8.5 Hz, 4H), 7.92 (s, 4H), 7.64 (s, 4H), 7.53 (dd, *J* = 16.0, 8.6 Hz, 8H), 7.44 – 7.36 (m, 4H), 1.84 (s, 12H) ppm.

**¹³C-NMR** (CD$_2$Cl$_2$, 75 MHz): δ = 140.7, 140.6, 139.1, 137.6, 136.1, 132.5, 130.4, 129.4, 128.2, 127.1, 126.8, 126.6, 126.5, 126.1, 20.71.

**HR-MALDI-TOF** (matrix: dithranol): calc. for [M]$^+$: 638.2973, found for [M]$^+$: 638.2932 (deviation: 6.4 ppm).

**HR-APCI-MS** (positive mode): calc. for [M+H]$^+$: 639.3052, found for [M+H]$^+$: 639.3040 (deviation: 1.88 ppm).

**Elemental analysis C$_{50}$H$_{38}$:** calc. for C: 94.00, H: 6.00; found for C: 91.95, H: 6.30 %.

**Melting point**: > 340 °C.

## 4. High-resolution mass spectra (MALDI-TOF, HR-ESI-MS, HR-APCI-MS)

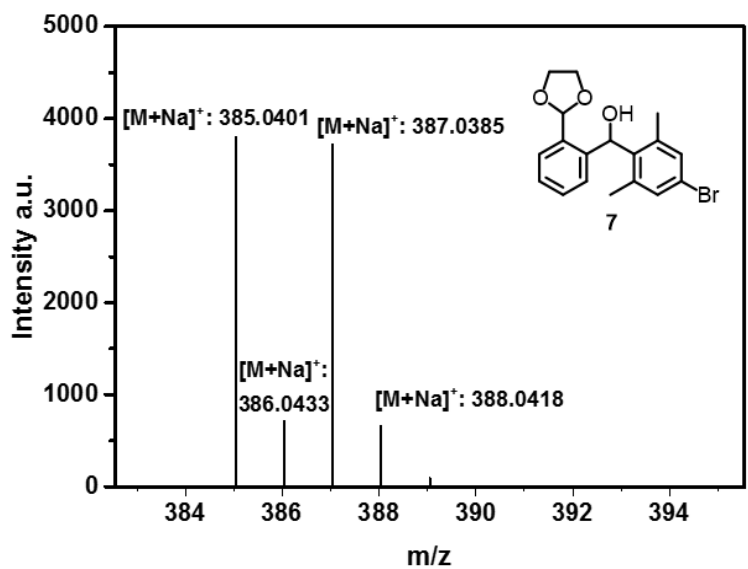

**Figure S4.** Liquid-state HR-ESI-MS (positive mode) of compound **7**.

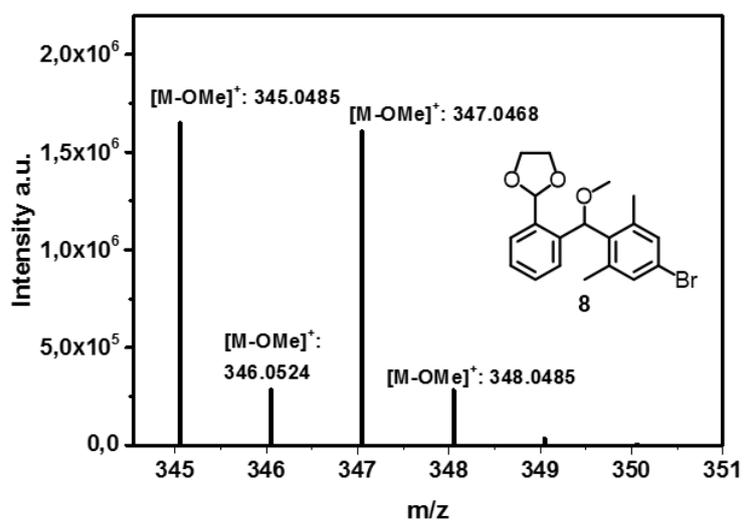

**Figure S5.** Liquid-state HR-APCI-MS (positive mode) of compound **8**.

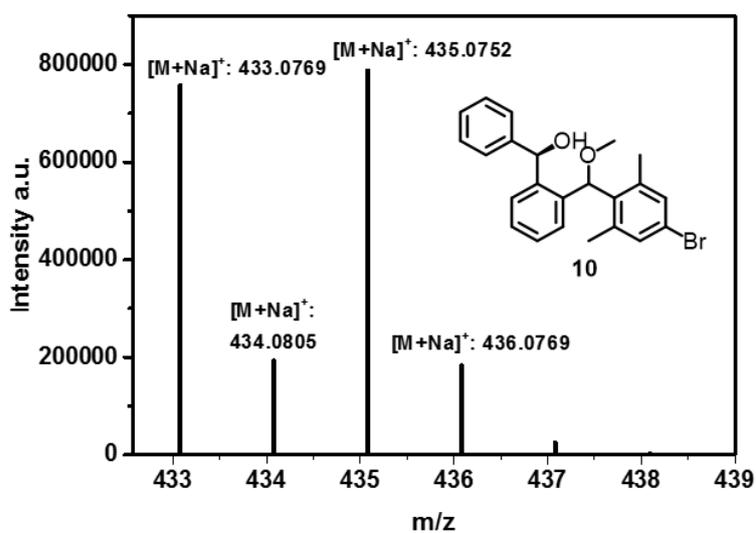

**Figure S6.** Liquid-state HR-ESI-MS (positive mode) of compound **10**.

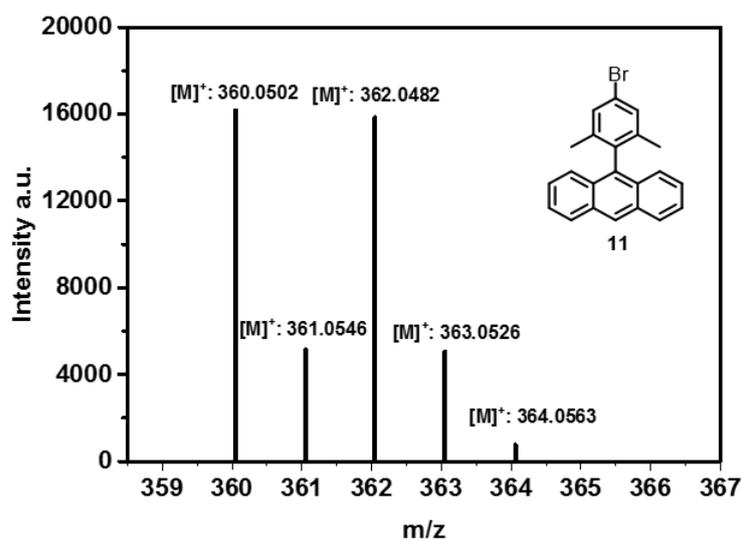

**Figure S7.** Liquid-state HR-ESI-MS (positive mode) of compound **11**.

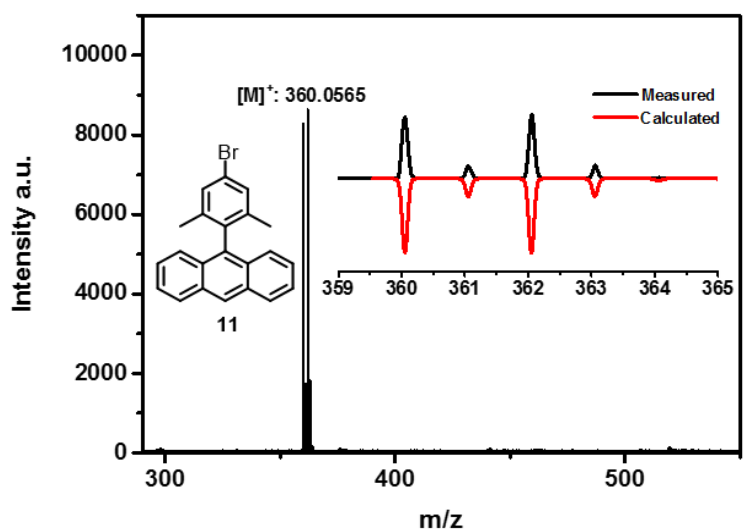

**Figure S8.** Liquid-state HR-MALDI-TOF (positive mode) of compound **11** (matrix: dithranol).

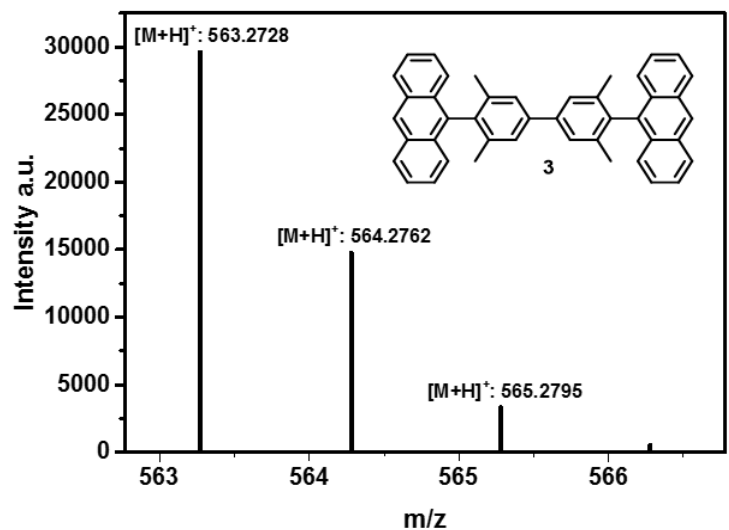

**Figure S9.** Liquid-state HR-APCI-MS (positive mode) of compound **3**.

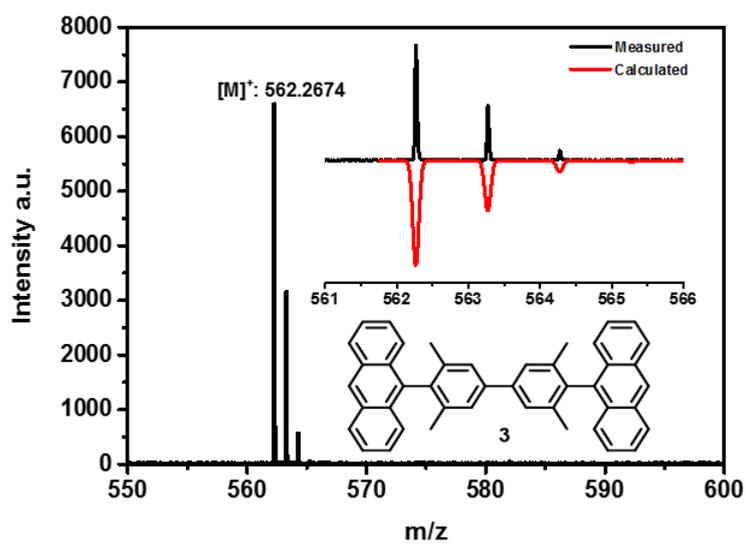

**Figure S10.** Liquid-state HR-MALDI-TOF (positive mode) of compound **3** (matrix: DCTB).

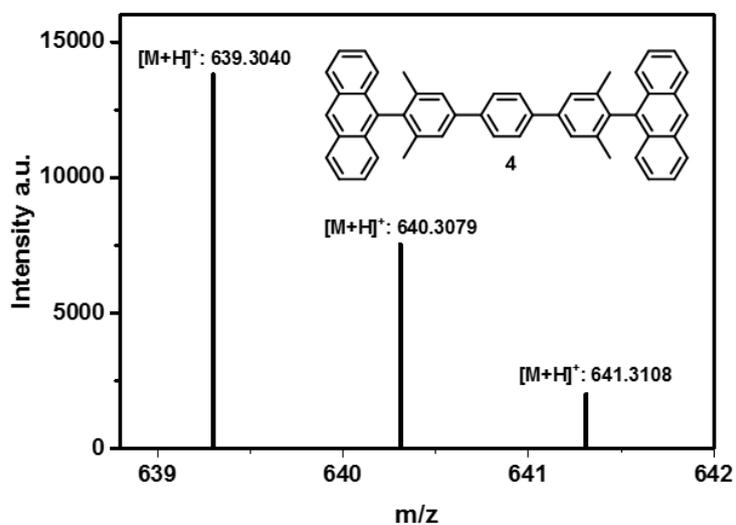

**Figure S11.** Liquid-state HR-APCI-MS (positive mode) of compound **4**.

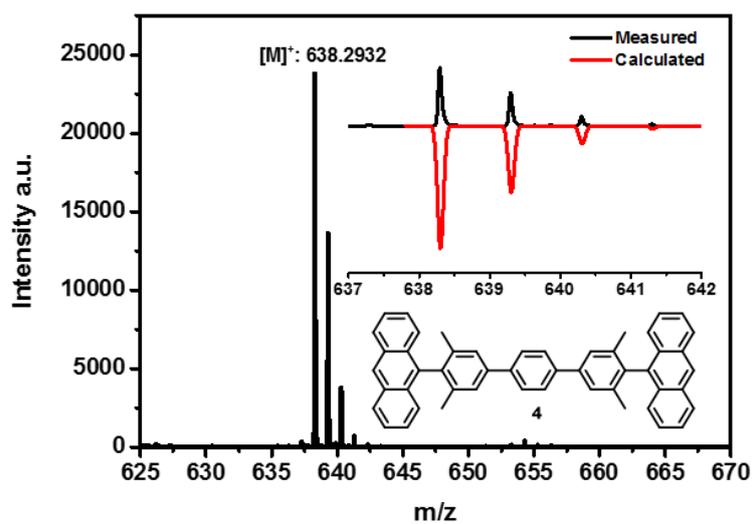

**Figure S12.** Liquid-state HR-MALDI-TOF (positive mode) of compound **4** (matrix: dithranol).

## 5. NMR Characterization ($^1$H-, $^{13}$C- and 2D-NMR)

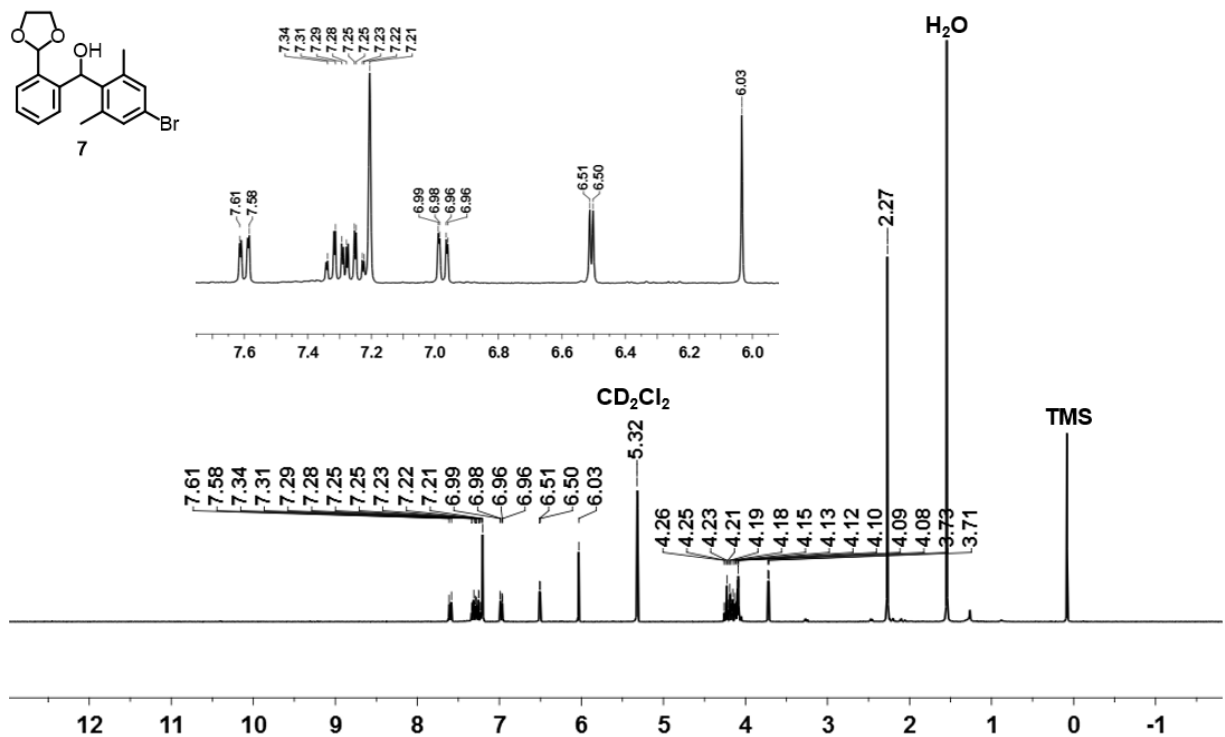

**Figure S13.** Liquid-state $^1$H-NMR spectrum of compound **7** measured in CD$_2$Cl$_2$ at room temperature. Frequency: 300 MHz.

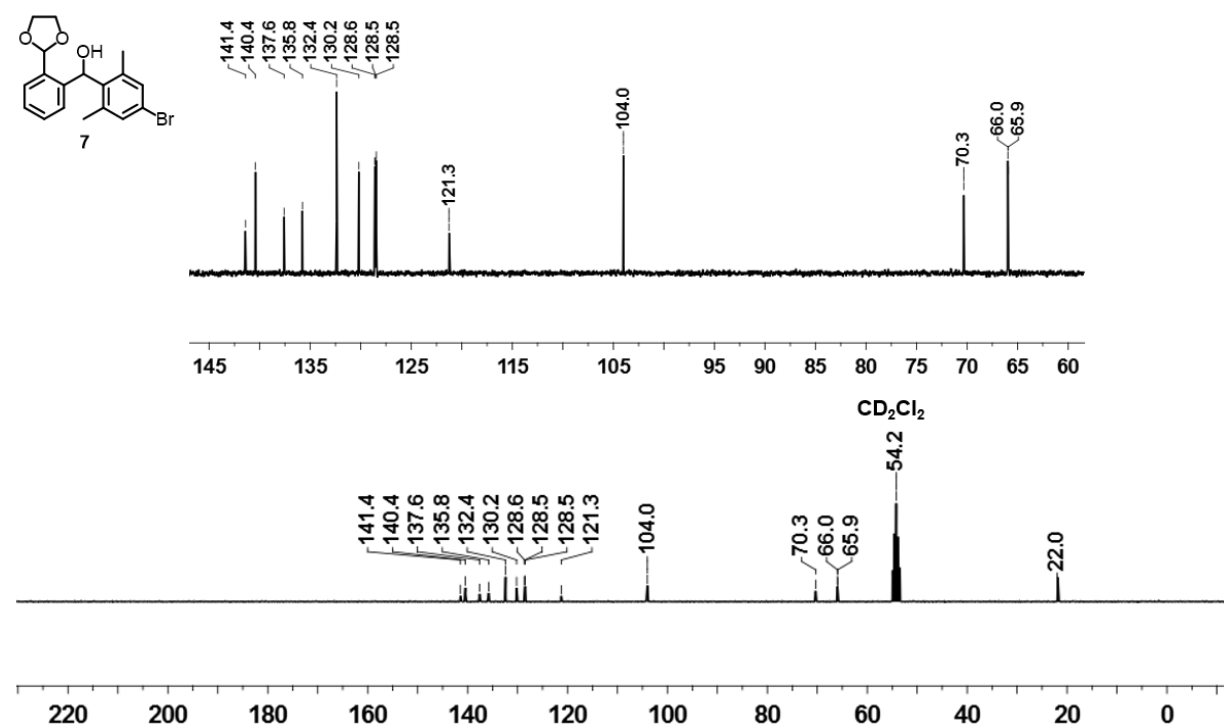

**Figure S14.** Liquid-state $^{13}$C-NMR spectrum of compound **7** measured in CD$_2$Cl$_2$ at room temperature. Frequency: 75 MHz.

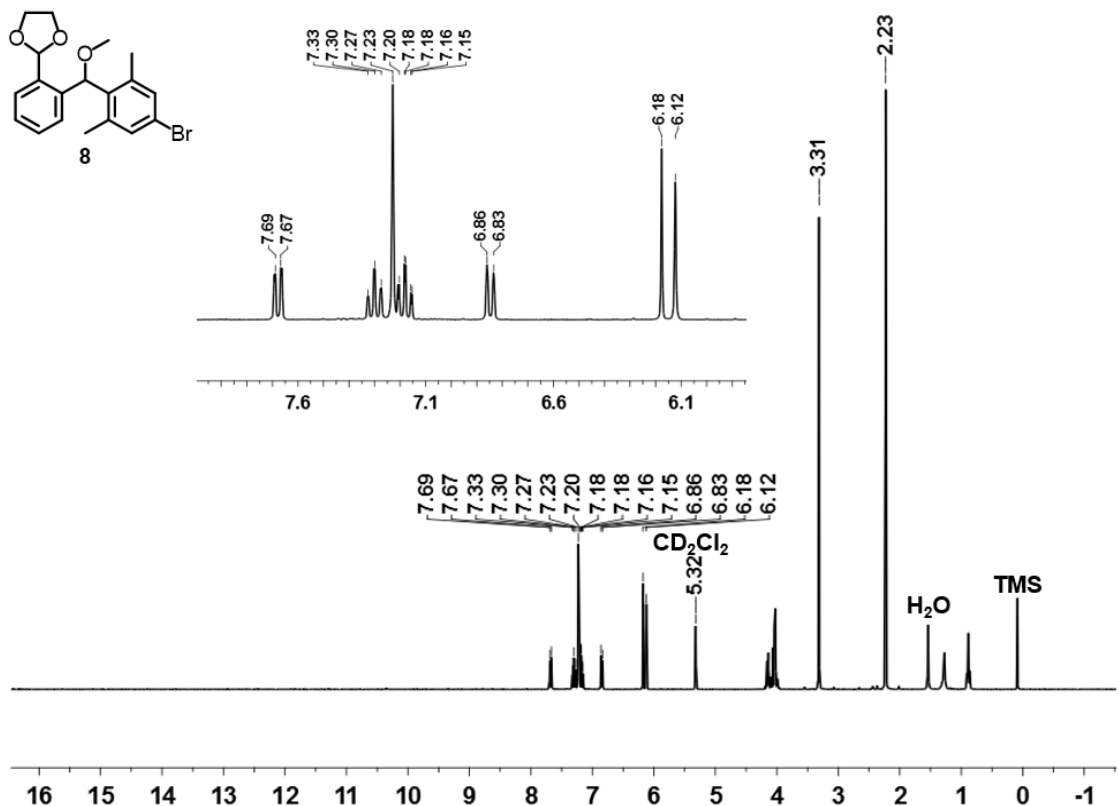

**Figure S15.** Liquid-state ¹H-NMR spectrum of compound **8** measured in CD$_2$Cl$_2$ at room temperature. Frequency: 300 MHz.

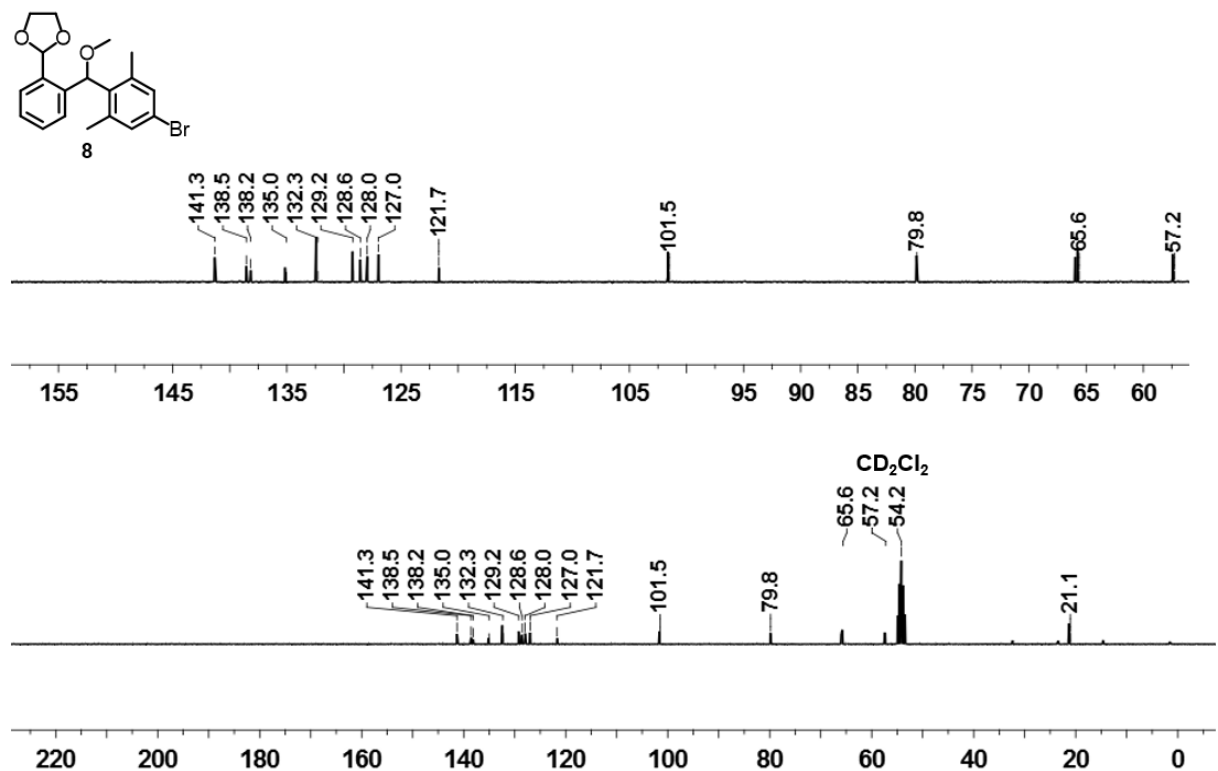

**Figure S16.** Liquid-state ¹³C-NMR spectrum of compound **8** measured in CD$_2$Cl$_2$ at room temperature. Frequency: 75 MHz.

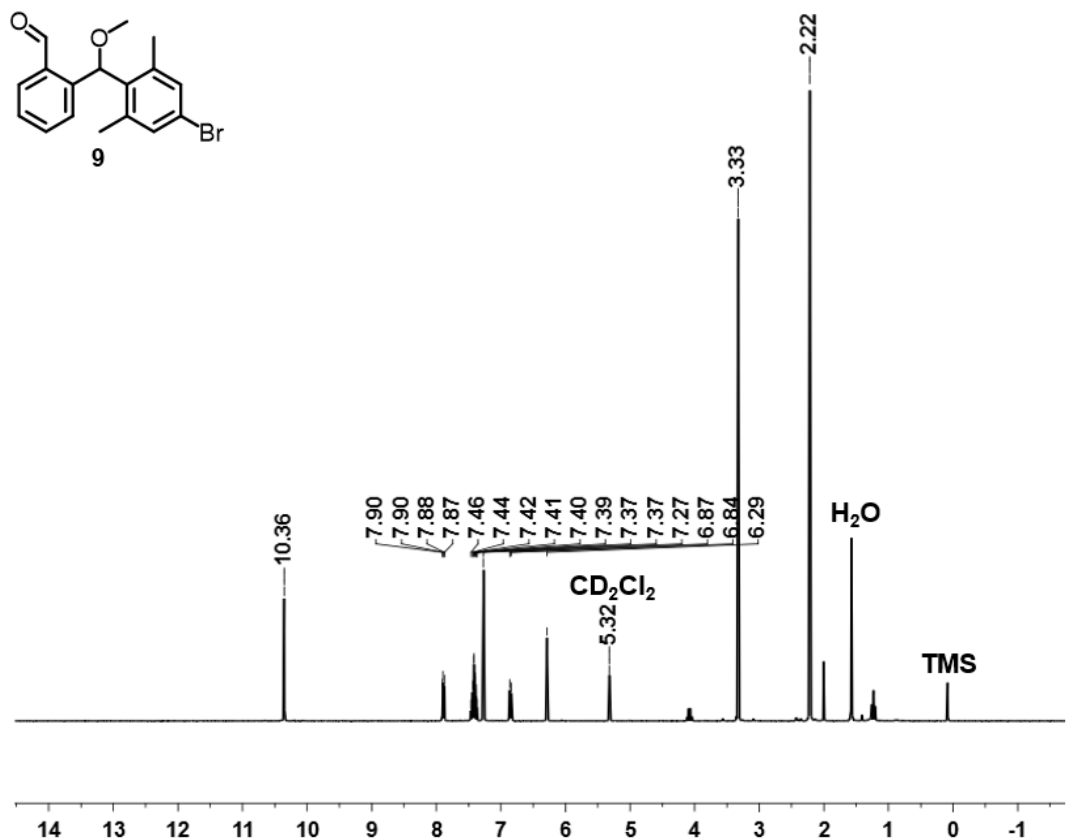

**Figure S17.** Liquid-state ¹H-NMR spectrum of compound **9** measured in CD$_2$Cl$_2$ at room temperature. Frequency: 300 MHz.

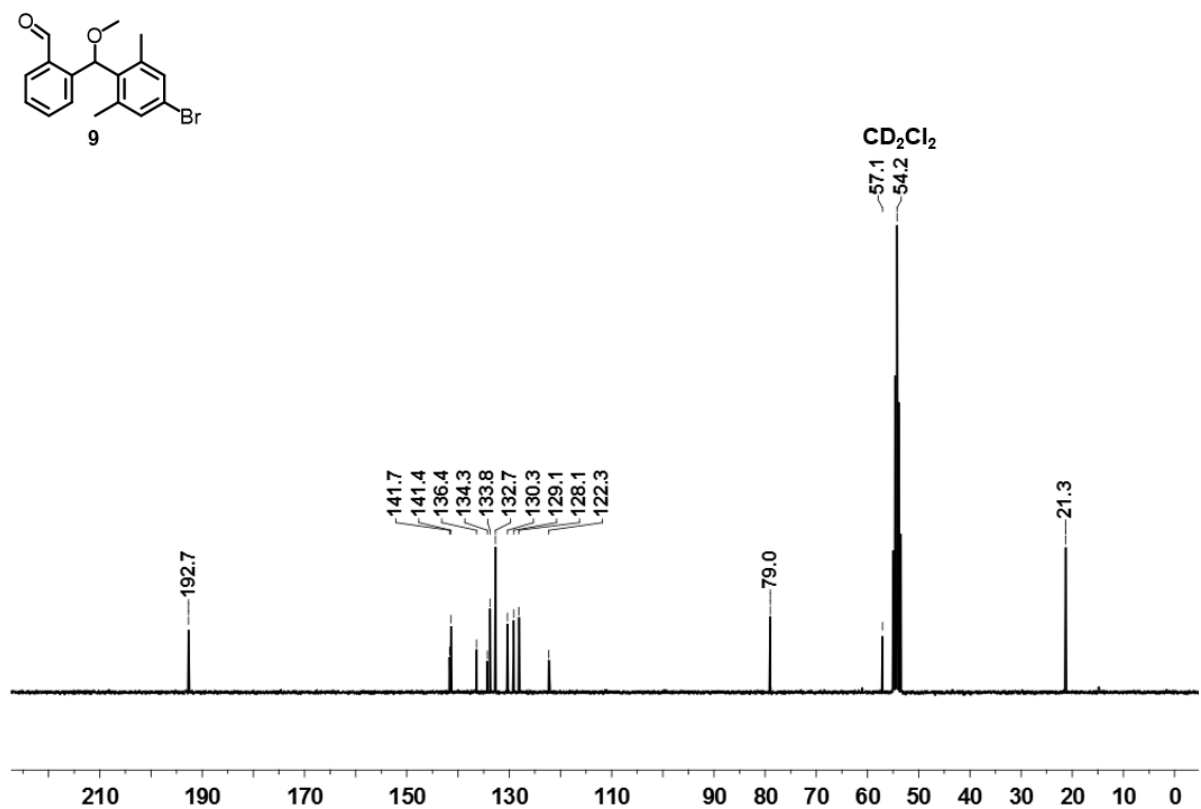

**Figure S18.** Liquid-state ¹³C-NMR spectrum of compound **9** measured in CD$_2$Cl$_2$ at room temperature. Frequency: 75 MHz.

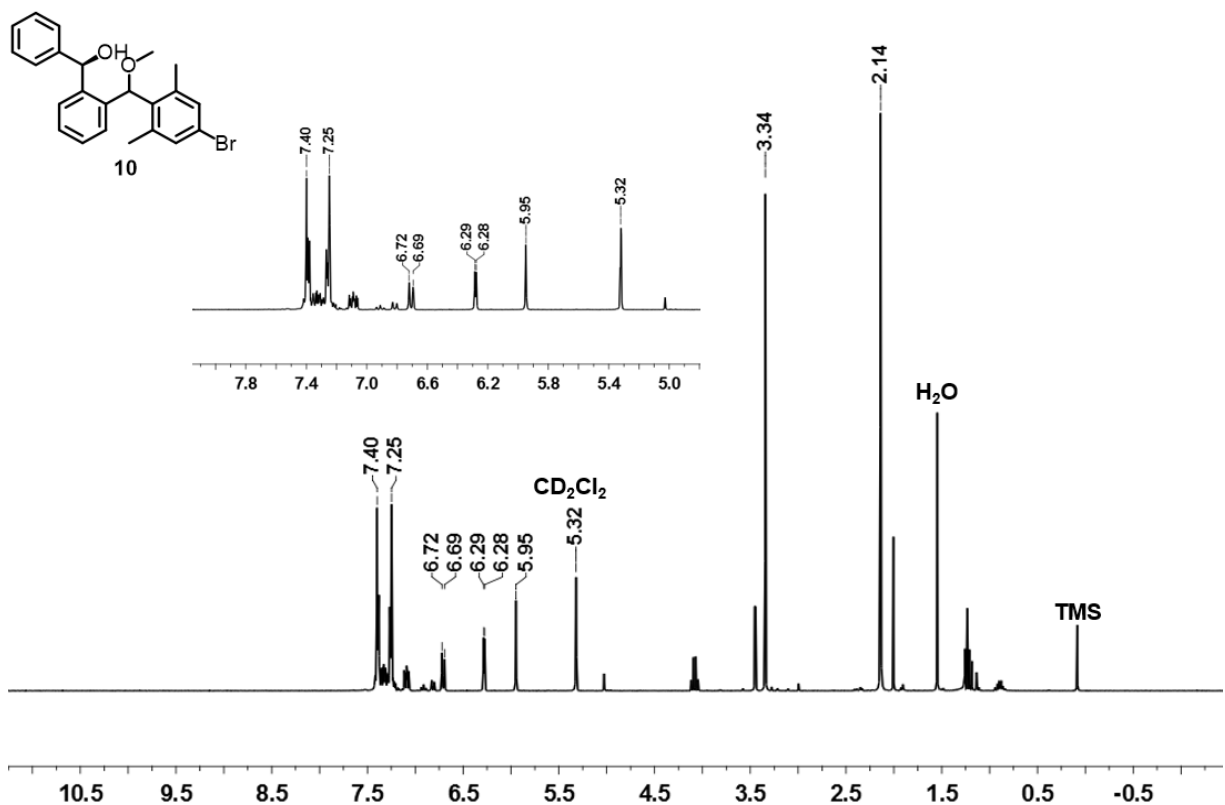

**Figure S19.** Liquid-state $^1$H-NMR spectrum of compound **10** measured in CD$_2$Cl$_2$ at room temperature. Frequency: 300 MHz.

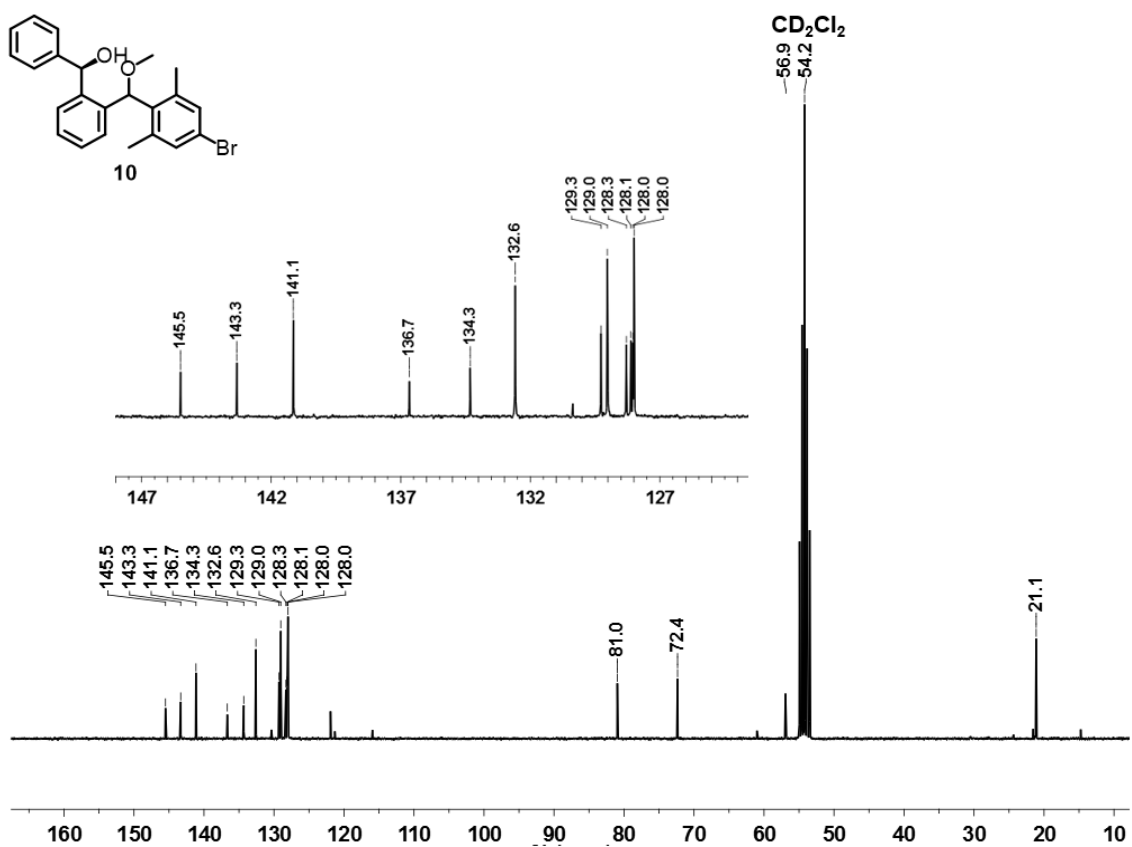

**Figure S20.** Liquid-state $^{13}$C-NMR spectrum of compound **10** measured in CD$_2$Cl$_2$ at room temperature. Frequency: 75 MHz.

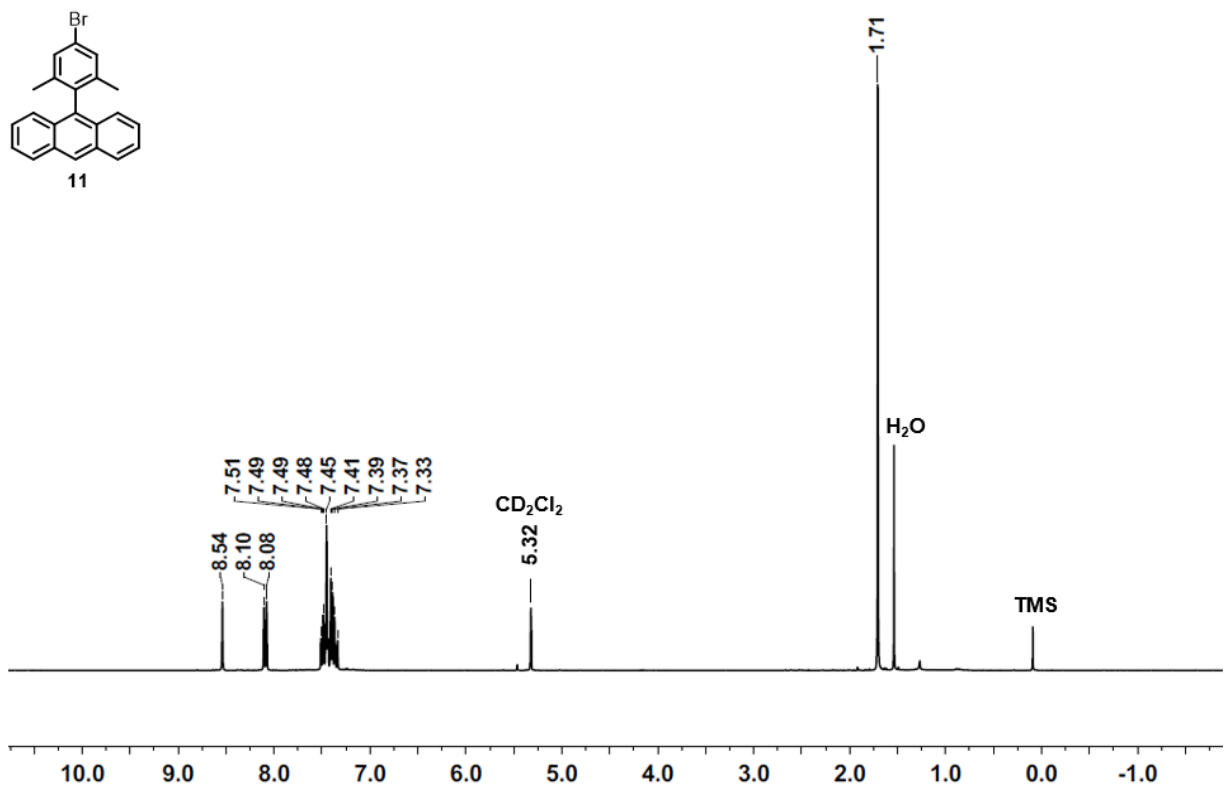

Figure S21. Liquid-state ¹H-NMR spectrum of compound 11 measured in CD$_2$Cl$_2$ at room temperature. Frequency: 300 MHz.

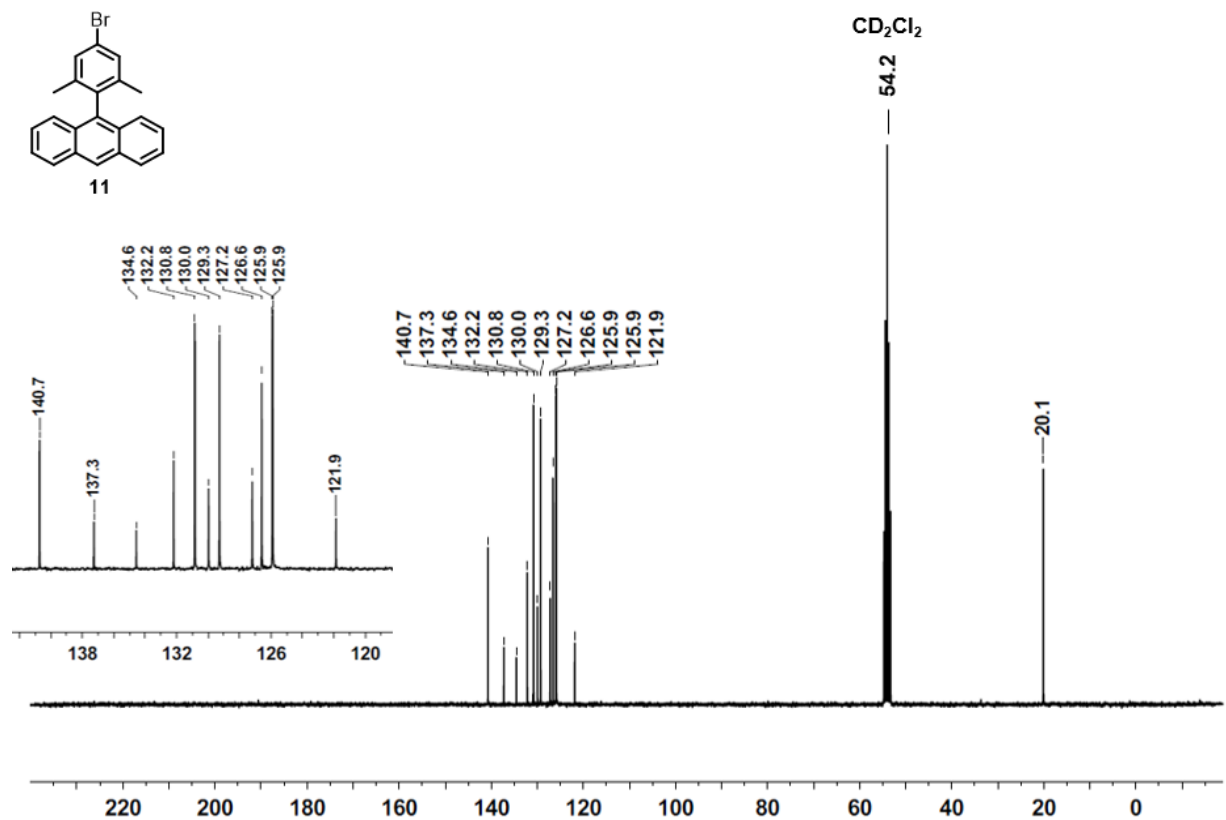

Figure S22. Liquid-state ¹³C-NMR spectrum of compound 11 measured in CD$_2$Cl$_2$ at room temperature. Frequency: 75 MHz.

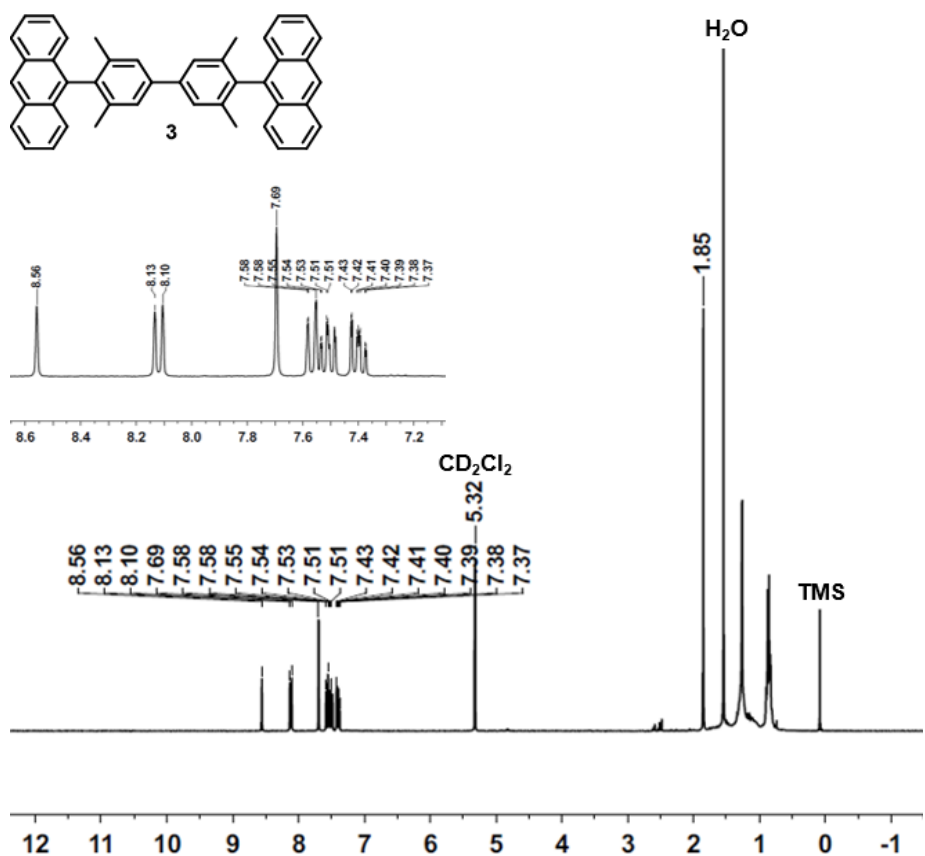

**Figure S23.** Liquid-state ¹H-NMR spectrum of compound **3** measured in CD$_2$Cl$_2$ at room temperature. Frequency: 300 MHz.

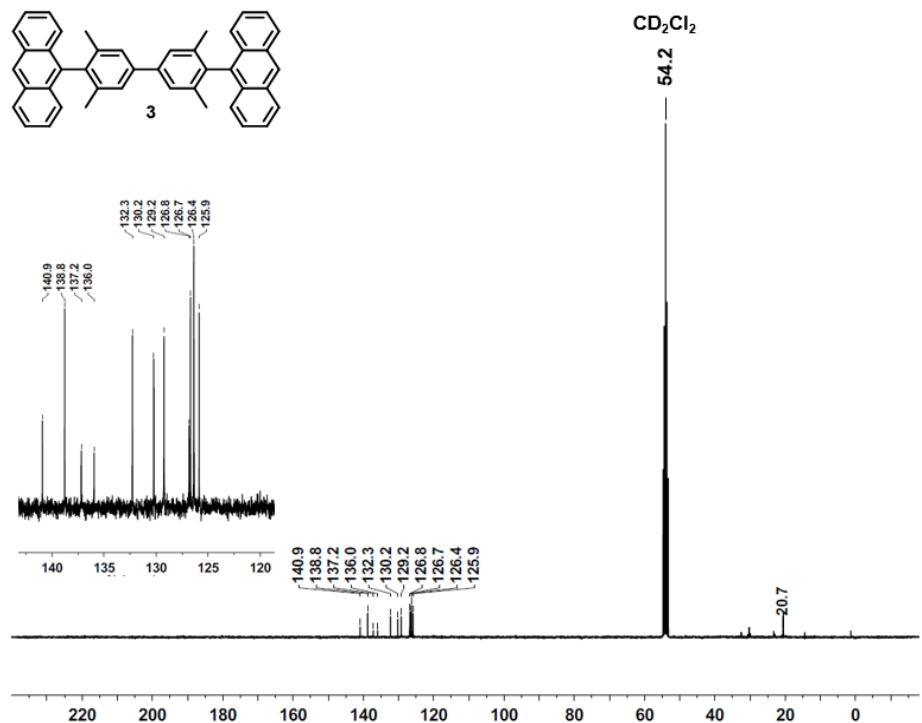

**Figure S24.** Liquid-state ¹³C-NMR spectrum of compound **3** measured in CD$_2$Cl$_2$ at room temperature. Frequency: 75 MHz.

S22

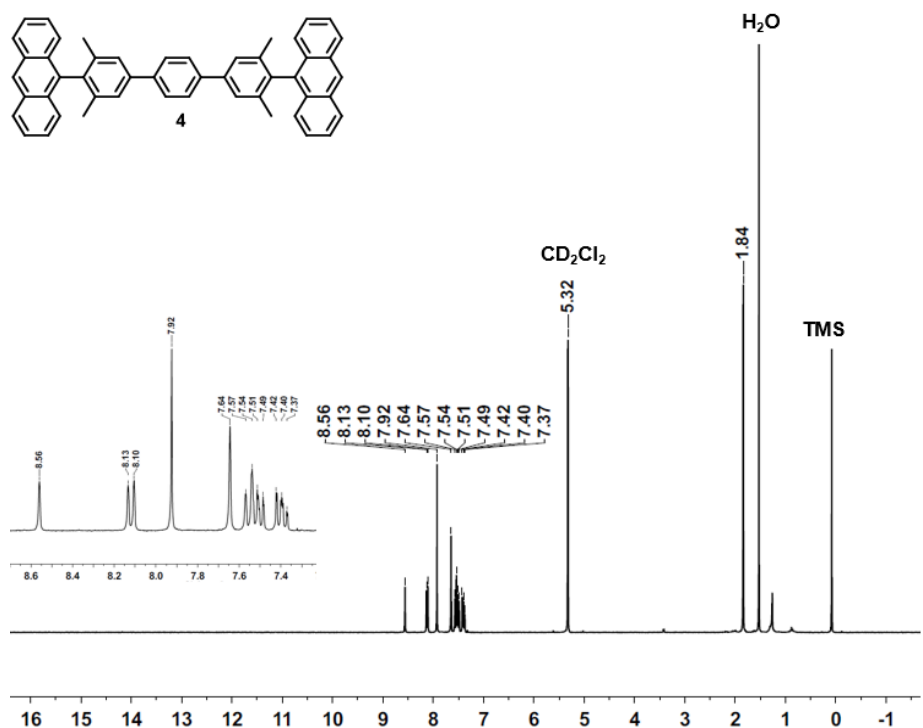

**Figure S25.** Liquid-state ¹H-NMR spectrum of compound **4** measured in CD$_2$Cl$_2$ at room temperature. Frequency: 300 MHz.

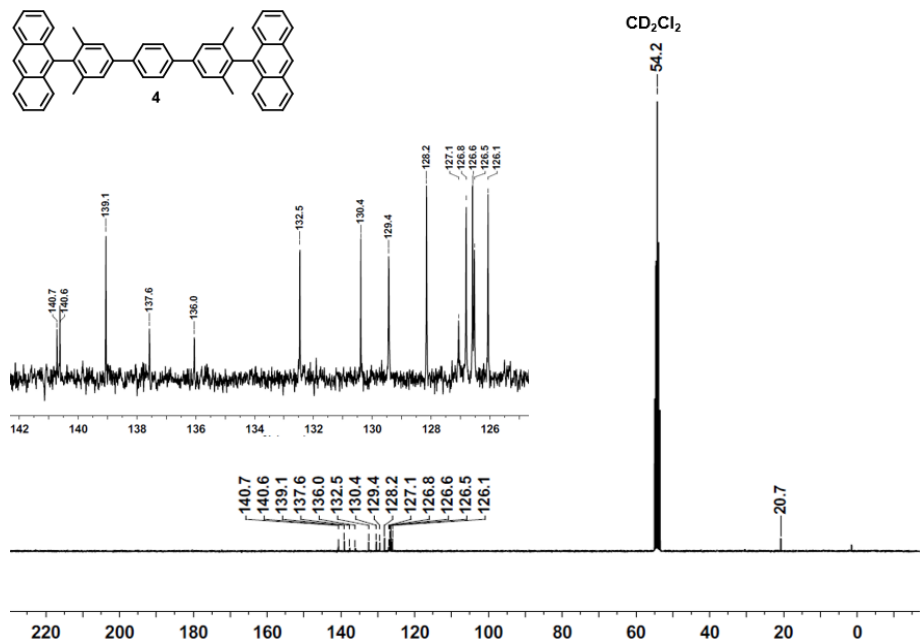

**Figure S26.** Liquid-state ¹³C-NMR spectrum of compound **4** measured in CD$_2$Cl$_2$ at room temperature. Frequency: 75 MHz.

## 6. References

[1] M. Ternes, *New J. Phys.* **2015**, *17*, 063016.

[2] I. Horcas, R. Fernández, J. M. Gómez-Rodríguez, J. Colchero, J. Gómez-Herrero, A. M. Baro, *Review of Scientific Instruments* **2007**, *78*, 013705.

[3] R. Ortiz, R. A. Boto, N. García-Martínez, J. C. Sancho-García, M. Melle-Franco, J. Fernández-Rossier, *Nano Lett.* **2019**, *19*, 5991–5997.

[4] R. Liang, T. Ma, S. Zhu, *Org. Lett.* **2014**, *16*, 4412–4415.